\documentclass[onecolumn, nobibnotes,prd,superscriptaddress,showpacs]{revtex4}
\usepackage{hyperref}
\usepackage{url}
\usepackage{graphicx}
\usepackage{dcolumn}
\usepackage{bm}
\usepackage{natbib}
\usepackage{amsmath}
\usepackage[usenames]{color}

\newcommand{\be}{\begin{equation}}
\newcommand{\ee}{\end{equation}}
\newcommand{\bea}{\begin{eqnarray}}
\newcommand{\eea}{\end{eqnarray}}

\newcommand{\WMAP}{{\slshape WMAP~}}

\newcommand{\s}{{\rm ~s}}

\newcommand{\g}{{\rm ~g}}
\newcommand{\cm}{{\rm ~cm}}
\newcommand{\km}{{\rm ~km}}

\newcommand{\MeV}{{\rm ~MeV}}
\newcommand{\mev}{{\rm ~MeV}}

\newcommand{\gev}{{\rm ~GeV}}
\newcommand{\TeV}{{\rm ~TeV}}
\newcommand{\epm}{{\ensuremath{e^+ e^-\;}}}

\newcommand{\kpc}{{\rm ~kpc}}

\definecolor{darkgreen}{rgb}{0,0.4,0} 

\begin{document}

\title{Sommerfeld-Enhanced Annihilation in Dark Matter Substructure: \\
Consequences for Constraints on Cosmic-Ray Excesses}

\author{Tracy R. Slatyer}
\email{tslatyer@ias.edu}
\affiliation{School of Natural Sciences, Institute for Advanced Study, Princeton, NJ 08540, USA}

\author{Natalia Toro}
\email{ntoro@perimeterinstitute.ca}
\affiliation{Perimeter Institute, Waterloo ON, Canada N2L 2Y5}
\affiliation{School of Natural Sciences, Institute for Advanced Study, Princeton, NJ 08540, USA}

\author{Neal Weiner}
\email{neal.weiner@nyu.edu}
\affiliation{Center for Cosmology and Particle Physics, Department of Physics, New York University, New York, NY 10003, USA}
\affiliation{School of Natural Sciences, Institute for Advanced Study, Princeton, NJ 08540, USA}

\begin{abstract}
In models of dark matter (DM) with Sommerfeld-enhanced annihilation, where the annihilation rate scales as the inverse velocity, $N$-body simulations of DM structure formation suggest that the local annihilation signal may be dominated by small, dense, cold subhalos. This contrasts with the usual assumption of a signal originating from the smooth DM halo, with much higher velocity dispersion. Accounting for local substructure modifies the  parameter space for which Sommerfeld-enhanced annihilating DM can explain the PAMELA and \emph{Fermi} excesses. Limits from the inner galaxy and the cosmic microwave background are weakened, without introducing new tension with substructure-dependent limits, such as from dwarf galaxies or isotropic gamma-ray studies. With substructure, previously excluded parameter regions with mediators of mass $\sim$ 1-200 MeV are now easily allowed. For $\mathcal{O}$(MeV) mediators, subhalos in a specific range of host halo masses may be evaporated, further suppressing diffuse signals without affecting substructure in the Milky Way.
\end{abstract}


\maketitle

\section{Introduction}
Interest in dark matter (DM) annihilation has been boosted in recent years as a consequence of a number of results from cosmic ray (CR) experiments. The PAMELA finding \cite{Adriani:2008zr} of a rise in the positron fraction at high ($\sim 10-100 \gev$) energies coupled with harder than expected $e^+ + e^-$ spectra from \emph{Fermi} \cite{Abdo:2009zk, Ackermann:2010ij} and ATIC \cite{aticlatest, Panov:2011zw} point to the existence of a new, primary source of high energy $e^+e^-$.  A more recent \emph{Fermi} measurement of the positron fraction at $20-200 \gev$ \cite{FermiLAT:2011ab} confirms the rise reported by PAMELA.

Attempts to explain the new $e^+ + e^-$ component of cosmic rays as DM annihilation products are immediately confronted by several difficulties: the $e^+$ and $e^-$ fluxes implied by the high-energy excesses are $\mathcal{O}(100-1000)$ times larger than expected from annihilation of a $\sim$ TeV-mass thermal relic, no corresponding excess is seen in the antiproton signal, and the positron spectrum is too hard to arise from DM annihilation into weak gauge bosons or hadrons. Models of TeV-scale DM coupled to light (MeV-GeV) force carriers  \cite{Finkbeiner:2007kk,ArkaniHamed:2008qn, Pospelov:2008jd} seek to address all three issues: kinematically forbidding the antiprotons, producing boosted positrons (giving rise to hard spectra), and allowing Sommerfeld enhancement of the annihilation cross section at low velocities \footnote{The Sommerfeld enhancement was first discussed in the context of DM by \cite{Hisano:2003ec, Hisano:2004ds}.}. Such models can explain the CR excesses while still yielding the appropriate thermal relic abundance, and without appealing to a local over-density of DM \cite{Finkbeiner:2010sm}.

\begin{figure*}
\includegraphics[width=0.45\textwidth]{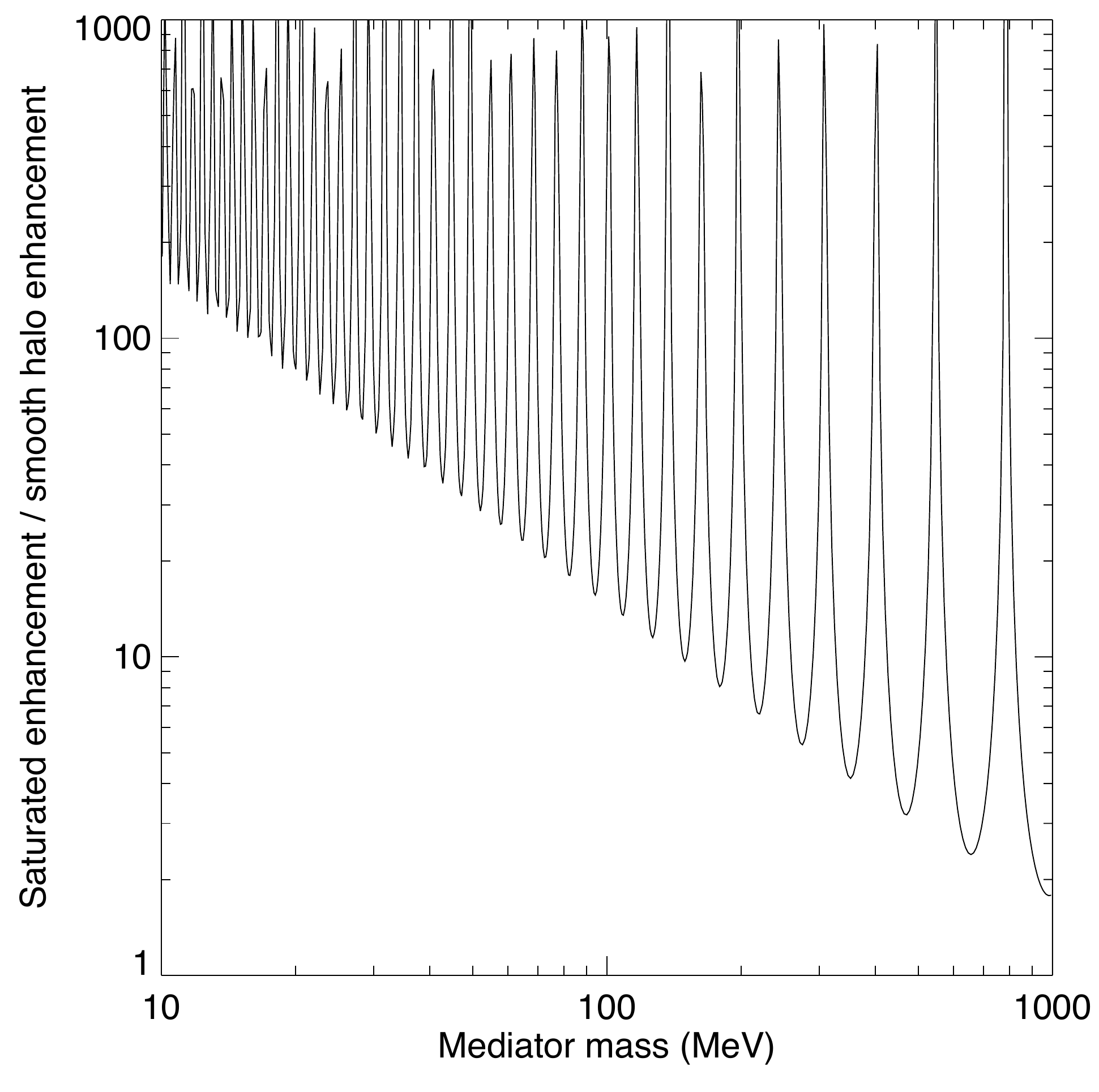}
\caption{The ratio of the saturated Sommerfeld enhancement to the smooth halo enhancement as a function of force carrier mass, assuming a local 1D velocity dispersion of $\sigma = 150$ km/s and a DM mass of 1.2 TeV.}
\label{fig:satlocalratio}
\end{figure*}

Most studies of such light-mediator models have assumed a smooth, Maxwellian halo in the local (within $\sim 1$ kpc) neighborhood, where the positrons and electrons observed by PAMELA and \emph{Fermi} are thought to originate.   But analytical arguments and $N$-body simulations both suggest a richer halo substructure, with a non-negligible fraction of DM residing in self-bound subhalos as small as $10^{-6} M_\odot$, as well as tidal streams and other structures.  Neglecting these subhalos would at first appear to be a well-justified approximation;  $N$-body simulations suggest that they contribute only an $\mathcal{O}(1)$ factor to the density-squared integral, and hence to the annihilation rate for conventional Weakly Interacting Massive Particles (WIMPs). However, bound subhalos typically have velocity dispersions much smaller than the $\sim$150 km/s of the smooth halo.  As Sommerfeld enhancement confers an additional $1/v$-$1/v^2$ scaling to the annihilation cross section (down to some saturation velocity, below which $\langle \sigma v \rangle$ is increased by a constant \emph{saturated enhancement}), even $\mathcal{O}(1)$ contributions to the density-squared integral can change the predicted Sommerfeld-enhanced DM annihilation rate by an order of magnitude or more.  As such, annihilation in local substructure can overwhelmingly dominate the smooth halo signal in models with Sommerfeld-enhanced annihilation (related effects have been studied in \cite{Lattanzi:2008qa, Kuhlen:2009is, Kuhlen:2009kx, Bovy:2009zs,Yuan:2009bb, Vincent:2010kv, Kistler:2009xf, Kamionkowski:2010mi}).    We show in Fig. \ref{fig:satlocalratio} the size of the saturated Sommerfeld enhancement relative to the Sommerfeld enhancement in the smooth halo. For heavier mediator masses, the difference is only a factor of a few (except on resonances), but for mediator masses $\lesssim 100 \mev$ the saturated enhancement is generically a factor of 10-100 times larger than the smooth-halo boost.

Many constraints on light-mediator models have been obtained under this assumption that the smooth halo dominates the local DM annihilation signal.   In other words, model parameters for which the smooth halo annihilation signal reproduces the results of PAMELA and \emph{Fermi} are used to normalize signals in constraining channels where no or little excess is seen.  Accounting for a substructure contribution to the local signal significantly changes the model parameters at which these  constraints should be studied, particularly for mediator masses in the $\sim 1-200$ MeV range where the local signal is naturally dominated by Sommerfeld-enhanced substructure.  As we shall see, a wide range of constraints become dramatically weaker in this case, specifically:
\begin{itemize}
\item Bounds on the cosmological DM annihilation rate between recombination and reionization, from the anisotropy power spectrum of the cosmic microwave background (CMB),
\item Constraints on light mediators from DM self-interactions,
\item Tensions between the relic abundance and the large local signal,
\item Non-observation of inverse Compton scattering (ICS) and final-state radiation (FSR) signals from annihilation in the Galactic center.
\end{itemize}
The parameter space with force carrier masses $\sim 1-200$ MeV, which is effectively ruled out in the smooth-halo-dominated case, is re-opened in the presence of an $\mathcal{O}(1)$ substructure contribution to the DM density-squared. This is an important point. 
Terrestrial fixed-target experiments and low-energy colliders can probe direct production of light
hidden-sector gauge bosons \cite{Pospelov:2008zw}.  Of these, collider searches \cite{Batell:2009yf,Essig:2009nc,Reece:2009un,:2009pw} can cover the widest range of gauge boson masses, while fixed-target experiments are sensitive to the widest range of couplings, but at lower masses \cite{Reece:2009un,Bjorken:2009mm,Batell:2009di}. Most fixed-target results \cite{Bergsma:1985qz,Riordan:1987aw,Bjorken:1988as,Bross:1989mp,Andreas:2010tp,Archilli:2011nh,Merkel:2011ze,  Abrahamyan:2011gv} and recent proposals \cite{Proposal,Essig:2010xa,HPS,Wojtsekhowski:2009vz,Freytsis:2009bh, HIPS} have the greatest reach at masses below $\sim 200$ MeV, the region where substructure effects can be most dramatic.  Understanding the role of substructure is essential to clarify what regions of parameter space have astrophysical motivation.

In this note, we will explore the added parameter space that is opened at light ($\sim$ 1-200 MeV) mediator masses when substructure is taken into account. In Sec. \ref{sec:constnosub}, we outline the key features of the Sommerfeld enhancement and simple models that illustrate them, and then review previously studied constraints on the relevant parameter space from bounds on DM self-annihilation and self-interaction. 
We do \emph{not} initially consider constraint channels that depend on the amount or distribution of substructure, deferring that discussion for later.  Sec. \ref{sec:effectsofsub} presents a simple parameterization of the boost to the annihilation rate in the presence of both local substructure and Sommerfeld enhancement, and shows how the interplay of these effects relaxes the constraints of Sec. \ref{sec:constnosub}. We study the parameter space favored to fit the PAMELA and \emph{Fermi} cosmic-ray signals as a function of the amount of substructure and the range of the DM self-interaction, the effect of the various constraints on this parameter space, and the effect of including substructure on the maximum possible annihilation rate (consistent with all constraints). In Sec. \ref{sec:constwsub}, we review and address constraints that depend on the distribution of substructure, both in the Milky Way (MW) and on cosmological scales.  These include limits from diffuse $\gamma$ rays, arising from substructure in the outer MW and in other halos, and limits from the inner MW, which depend upon the extent to which substructure persists in the inner galaxy.

\section{A Review of Relevant Constraints on Dark Matter Annihilation and Self-Interaction}
\label{sec:constnosub}

\subsection{Modeling the Sommerfeld enhancement}
\label{sec:lagrangian}

Here we briefly review the physics of Sommerfeld enhancement and summarize a simple concrete model for the enhancement, discussed in detail in \cite{Finkbeiner:2010sm}. For illustration, consider the long-range self-interaction due to a Yukawa potential. More complicated potentials are certainly possible, but this simplest scenario captures several key features: the $1/v$ scaling of the enhancement, the presence of resonances where the scaling becomes $1/v^2$, and the saturation of the enhancement due to the finite range of the potential. We will denote the force carrier by $\phi$, with mass $m_\phi$, and the DM by $\chi$, with mass $m_\chi$. The enhancement factor -- which physically is simply due to the focusing effect of the attractive interaction -- can be derived by solving the Schr\"odinger equation with a Yukawa potential,
\begin{equation} \frac{1}{m_{\chi}} \psi''(r) - V(r) \psi(r) = -m_{\chi} v^2\psi(r), \quad V(r) = -\frac{\alpha_D}{r} e^{- m_{\phi} r},  \label{schrodinger} \end{equation}
where $v$ is the velocity of each particle in the center-of-mass frame (here we use units where $\hbar=c=1$), and $\alpha_D$ describes the coupling of the DM to the force carrier.

Such a potential can be generated by coupling a scalar $\phi$ to Majorana fermion DM through a Yukawa term in the Lagrangian,
\begin{equation} \mathcal{L}_\mathrm{int} = -\frac{1}{2} \lambda \phi \chi \chi, \quad \alpha_D = \frac{\lambda^2}{4 \pi}. \end{equation}
It can also arise in models with a vector mediator: in the examples from \cite{Finkbeiner:2010sm}, the DM is a (pseudo-)Dirac fermion $\Psi$ charged under a new dark U(1), with the symmetry broken at the GeV scale by a dark Abelian Higgs field $h_D$. 

At energies below the symmetry breaking scale, the DM can be regarded as a pair of Majorana fermion species (which may or may not be degenerate in mass, depending on other operators in the theory). We can write $\Psi = (\chi, \eta^\dagger)$, where $\chi, \eta$ are mass-degenerate two-component spinors of opposite U(1)-charge, following the notation of \cite{Dreiner:2008tw}. If the DM field acquires a Majorana mass, breaking the mass degeneracy, the mass eigenstates $\chi_1, \chi_2$ are rotated 45$^\circ$ from the gauge eigenstates. Thus the part of the Lagrangian describing the interaction between the DM and the new force carrier is given by,
\begin{align} \mathcal{L}_\mathrm{int} & =  - g_D \bar{\Psi} \gamma^\mu \phi_\mu \Psi = - g_D \phi_\mu \left(\chi^\dagger \bar{\sigma}^\mu \chi - \eta^\dagger \bar{\sigma}^\mu \eta \right) = - i g_D \phi_\mu \left( \chi_1^\dagger \bar{\sigma}^\mu \chi_2 - \chi_2^\dagger \bar{\sigma}^\mu \chi_1 \right), \quad \mathrm{with} \, \alpha_D = \frac{g_D^2}{4 \pi}. \end{align}
Such a model will also contain other interactions, certainly between the dark Higgs and the gauge bosons, and possibly between the dark Higgs and the DM (the Yukawa coupling is suppressed by the small mass of the dark Higgs, but there may be higher-dimension operators with larger effects, in particular if one wishes to generate a Majorana mass term). These interactions can open additional annihilation channels. A complete discussion is given in \cite{Finkbeiner:2010sm}; in brief, most of these channels are suppressed at late times (although they may contribute to DM annihilation during freezeout), with the possible exception of the $s$-channel process $\chi_1 \chi_2 \rightarrow \phi h_D$ in the case where the mass eigenstates are degenerate. This additional unsuppressed channel experiences Sommerfeld enhancement as usual, but the exact SM final state (and hence the contribution to the cosmic-ray signal) will depend on the relative masses of $\phi$ and $h_D$ and any further light states in the dark sector.

For simplicity, let us restrict ourselves to the minimal case, where only the $\chi \chi \rightarrow \phi \phi$ annihilation channel contributes to the cosmic-ray signal, and the potential can be approximated as Yukawa. We do not claim that this represents the complete set of models with Sommerfeld-enhanced annihilation (and an additional example, with a splitting between the mass eigenstates, is discussed briefly in Appendix \ref{app:inelastic}, where the complete Lagrangian is also given). However, it adequately demonstrates the interplay of the various constraints.

The Sommerfeld enhancement due to such a Yukawa potential scales as $\pi \alpha_D/v$ in the regime where $v/c \gtrsim m_\phi/m_\chi$, where $m_\phi$ is the mass of the force carrier, $m_\chi$ is the mass of the DM, and $\alpha_D$ is the coupling of the DM to the force carrier. At $v/c \sim m_\phi/m_\chi$ the enhancement saturates due to the finite range of the force; a good estimate for the saturated enhancement is $12 \alpha_D m_\chi / m_\phi (1 - \cos \theta)$, where the angle $\theta \approx 2 \sqrt{6} \sqrt{\alpha_D m_\chi / m_\phi} $ describes the resonance structure \cite{Cassel:2009wt, Slatyer:2009vg}. Points where $\theta = 2 \pi n$ correspond to values of $\alpha_D m_\chi/m_\phi$ where the potential has a zero-energy bound state; close to these resonances, the enhancement instead scales as $\sim 1/v^2$ up to saturation. Thus, the saturated low-velocity enhancement exceeds the enhancement at intermediate velocities $v$ (as in the smooth local halo) by a ratio $\sim 4 m_\chi v / m_\phi (1 - \cos \theta) \ge 2 m_\chi v / m_\phi$.

\subsection{Constraints on the saturated enhancement}

Due to the rapid scaling of the Sommerfeld enhancement with velocity, strong constraints on the parameter space originate from systems where the typical DM velocity is very small (in particular, dwarf galaxies and the early universe). The potentially large ratio between the enhancement in the smooth local halo and the low-velocity saturated enhancement allows such limits to be translated into strong bounds on the local DM annihilation rate. In particular, such searches are sensitive to small mediator masses, due to the $1/m_\phi$ scaling of the ratio, and the regions around resonances, where this ratio is large as shown in Fig. \ref{fig:satlocalratio}.

In the absence of substructure, the strongest constraints in this category arise from measurements of the CMB\footnote{However, if the dwarf galaxies contain significant substructure, the bounds on the low-velocity annihilation cross section from dwarfs will be strengthened; we consider this bound in substructure-dominated scenarios in Sec. \ref{subsec:dwarfs}.}. DM annihilation during the epoch of recombination injects ionizing electrons and photons which broaden the last scattering surface and give rise to increased damping of temperature anisotropies, combined with enhanced polarization anisotropies \cite{Padmanabhan:2005es}. The annihilation cross section can therefore be constrained by high-precision measurements of the CMB. The typical velocity of WIMPs at $z \sim 1000$ is of order $v \sim 10^{-8} c$ \cite{Galli:2009zc}, so we expect the Sommerfeld enhancement to be saturated; the bound from \WMAP 5 can be summarized as $\langle \sigma v \rangle_\mathrm{sat} \lesssim (120/f) \left( m_\chi / 1 \mathrm{TeV} \right) 3 \times 10^{-26} \mathrm{cm}^3/\mathrm{s}$, where $f \sim 0.2-0.7$ is an efficiency factor depending on the annihilation final state (see \cite{Slatyer:2009yq} for details)\footnote{More recent analyses using \WMAP7 data have found limits lower by a factor of $\sim 2/3$ \cite{Galli:2011rz} and larger by a factor of $\sim 19/14$ \cite{Hutsi:2011vx}: the reason for this factor-of-2 discrepancy is not well understood, and for simplicity we use the \WMAP 5 limits for illustration in this work.}. This bound is typically only a factor of $\sim 2-3$ higher than the local annihilation rate required to fit the CR excesses; consequently, if the smooth halo dominates the local signal, mediator masses lighter than $\sim 200$ MeV can be ruled out, at least for simple Sommerfeld models \cite{Finkbeiner:2010sm}.

\subsection{Self-interaction limits}
The Sommerfeld enhancement is simply the effect of virtual force carrier exchanges prior to DM annihilation. The same force carriers that mediate the Sommerfeld enhancement will give rise to a long-range self-interaction for the DM, which can be constrained by studying the longevity and morphology of various astrophysical systems.

Buckley and Fox (\cite{Buckley:2009in} and references therein) identify seven classes of constraints, from observations of the Bullet Cluster, evaporation of galaxies and dwarf galaxies, the stability of elliptical cores in galaxy clusters, the growth rate of supermassive black holes, thermodynamics of galaxies, and the structure of dwarf galaxies.  Feng, Kaplinghat, and Yu independently examined the constraints from elliptic galaxies \cite{Feng:2009hw}.

Each of these constraints can be formulated in terms of a velocity-averaged \emph{transfer cross-section} at an appropriate characteristic velocity, which for a particle of mass $m_{\chi}$ in a Yukawa potential controlled by the mediator mass $m_\phi$ and coupling $\alpha_D$ is well approximated by \cite{PhysRevLett.90.225002},
\begin{equation}
\sigma_T \approx \left\{ \begin{array}{cc} \frac{4\pi}{m^2_\phi} \beta^2 \ln(1+\beta^{-1}),
& \beta<0.1, \\
\frac{8\pi}{m^2_\phi}
\beta^2/(1+1.5\beta^{1.65}),
& 0.1 \le \beta \le 1000, \\
\frac{\pi}{m^2_\phi}
\left( \ln\beta + 1 - \frac{1}{2} \ln^{-1} \beta \right)^2,
& \beta > 1000,
\end{array} \right.
\label{eq:sigmat}
\end{equation}
where $\beta = 2 \alpha_D m_\phi / (m_\chi v_\mathrm{rel}^2)$.
Limits on $\sigma_T$ in systems of different velocity dispersions can be formulated as upper limits on $\alpha_D$, as a function of $m_\phi$ and $m_\chi$, and compared to one another.   

The shape of dwarf galaxies presents a particularly strong potential constraint on Sommerfeld-enhanced models, because it depends on self-interaction at velocities as low as 10 km/s, where the scattering cross section from light mediators can be large. Transfer cross-sections  $\sigma/m_{\chi} \gtrsim 0.1 \cm^2/\g$ at velocity dispersions $v_0 \approx 10$ km/s are expected to cause significant departures in halo structure from cold DM models  (see e.g. \cite{Hannestad:2000bs}).  

Transfer cross-sections above this ``self-interaction threshold'' would significantly affect the structure of dwarf galaxies, though such effects may not be ruled out. Indeed, it has even been argued that such a velocity-dependent force can explain the origin of cores in dwarf galaxies \cite{Loeb:2010gj}.  Some literature has nonetheless treated this threshold as a limit, and we include it in our numerical results.  A more conservative upper limit is  $\sigma/m_{\chi} \lesssim 5.6 \cm^2/\g$, the threshold for collapse of the dwarf halo's core in less than a Hubble time \cite{Hannestad:2000bs} (we thank M. Kuhlen for bringing this point to our attention).  In Sommerfeld-enhanced models, the latter limit is never reached in parameter regions permitted by the dwarf evaporation limit, discussed below.   The ``self-interaction threshold'' cross-section is reached even for tiny couplings at low mediator masses, but rapidly becomes irrelevant at high mediator masses.  For example, for $m_\chi = 1 \TeV$ and mediators masses below 7 MeV, $\sigma /m_{\chi} \lesssim 0.1 \cm^2/\g$ requires  $\alpha_D \lesssim 10^{-4}$, but for larger mediator masses, it requires
\begin{equation}
\alpha_D \lesssim 0.023 \times \left(\frac{20 \MeV}{m_\phi}\right)^{4.7}   \qquad (7 \lesssim m_\phi \lesssim 20 \MeV).
\end{equation}
Thus, self-interaction effects are typically not significant for Sommerfeld-enhanced explanations of the PAMELA and \emph{Fermi} excesses with $m_\phi > 20 \MeV$.
While this ``bound'' is much weaker than that arising from the CMB, we shall show that it becomes the leading constraint on light-mediator models with significant substructure.  

A weaker, but more robust constraint arises from the \emph{evaporation} of dwarf galaxies, which however depends on the velocity dispersion of the host galaxy.  The presence of dwarf galaxies in the MW implies a bound  $\sigma_T/m_{\chi} \lesssim 0.1 \cm^2/\g$ at velocities  $v_0 \approx 100$ km/s.  This bound permits $\alpha_D$ approximately 100 times larger than the self-interaction threshold ($\alpha_D\lesssim 0.01$ for $m_{\phi}<20$ MeV).  Even in the presence of significant substructure, this constraint is weaker than the one from the CMB. 

\subsection{Relic density limits}
\label{subsec:relicdensity}

There has recently been some debate over whether models with Sommerfeld-enhanced annihilation can provide a large enough cross section to fit the local CR data at all, while remaining consistent with the measured DM relic density: in other words, the upper limit on the DM--mediator coupling required to avoid over-depletion of the DM density has been said to rule out the entire parameter space in which such models can fit the CR excesses. Specifically, assuming the signal originates entirely from the smooth halo, \cite{Feng:2010zp} found that the maximal local enhancement was too low by a factor of $\sim 15$ to explain the CR signals, assuming a $2.35$ TeV DM candidate annihilating through a light force carrier solely into muons (providing a good spectral fit to the data), with a local halo density of $0.3$ GeV/cm$^3$. 

However, this claimed tension was based on a rather small region of parameter space, within a limited class of models. Using a more up-to-date local density estimate of $0.4$ GeV/cm$^3$, and specific models where the light force carrier was a vector and decayed to Standard Model (SM) states according to their charge (via kinetic mixing with the photon), \cite{Finkbeiner:2010sm} found that the discrepancy was a factor of $\sim 3$ or less (depending on the DM mass and final state) if the states in the dark-charged DM multiplet were taken to be degenerate, and that there was no discrepancy if a small splitting ($\sim 0.1-1$ MeV) between the states was permitted.

\subsection{Inner Galaxy limits}

Finally, limits on gamma-ray emission from around the center of the MW place constraints on DM annihilation which can be applied to models with Sommerfeld-enhanced annihilation, albeit with large astrophysical uncertainties. Such limits have been studied by e.g.  \cite{Bell:2008vx,Bertone:2008xr,Bergstrom:2008ag,Cirelli:2009vg,Pato:2009fn,Meade:2009iu,Cirelli:2009dv,Papucci:2009gd,Hutsi:2010ai}. 

The gamma-ray signal from DM annihilation has two components: (1) photons produced in the annihilation itself, either as FSR or from decays of neutral pions, and (2) starlight, infrared and CMB photons which are inverse Compton scattered to gamma-ray energies by high-energy $e^+ e^-$. We label these two components as ``FSR'' and ``ICS'' respectively. Both components depend on the DM density profile and the velocity profile in Sommerfeld-enhanced models; the ICS component tends to provide much stronger constraints than FSR alone for models fitting the PAMELA and \emph{Fermi} excesses, but such limits rely on an accurate model for cosmic ray propagation.

$N$-body simulations of cold DM structure formation predict a DM density profile with a pronounced peak (``cusp'') in the Galactic center. Observations of low-surface-brightness disk galaxies and dwarf galaxies indicate that such steep cusps are not present in those systems: instead, shallow ``cores'' in the density distribution are found (see e.g. \cite{2010AdAst2010E...5D} for a review and \cite{Oh:2010mc} for recent data). However, for the MW and similar galaxies, there is no observational evidence either for or against the presence of such a core.

The most recent conservative analyses of the gamma-ray signal from the inner MW \cite{Cirelli:2009dv,Papucci:2009gd,Hutsi:2010ai} indicate that models fitting the CR data remain allowed if the final state consists of electrons and/or muons, and the DM density profile possesses a core. For DM density profiles that instead possess a cusp with the properties favored by $N$-body simulations, there is severe tension between models fitting the CR data and gamma-ray limits; a recent analysis using less conservative assumptions finds tension even for a cored profile \cite{Zaharijas:2010ca}. These analyses assume that the local and inner Galaxy signals are both dominated by the smooth halo, and the annihilation cross section does not change with Galactocentric radius.

How these constraints change when Sommerfeld enhancement is included is not clear, simply because the velocity distribution of DM in the inner Galaxy is also not well known. While some authors have included models for the velocity dispersion motivated by $N$-body simulations (see e.g. \cite{Cirelli:2010nh, Navarro:2008kc}), the presence of baryons is expected to significantly affect the density and velocity profiles of DM in the inner Galaxy, and even the sign of the effect is not clear (see e.g. \cite{1986ApJ...301...27B,2004ApJ...616...16G, ElZant:2001re,ElZant:2003rp, RomanoDiaz:2008wz, Governato:2009bg, RomanoDiaz:2009yq,Abadi:2009ve,Pedrosa:2009rw,Tissera:2009cm}). For the ICS signal, which is essential to obtaining the strongest constraints on models which do not produce copious neutral pions, the magnetic field of the inner Galaxy plays an important role, and the presence of gamma-ray structures suggesting a possible large-scale high-energy outflow from the Galactic Center (GC) \cite{Su:2010qj} calls into question the usual steady-state modeling of CR propagation in this region of the sky. 

Due to these astrophysical uncertainties, we do not impose the inner Galaxy limits when determining our preferred regions of parameter space. Nonetheless, for non-cored DM density profiles the FSR signal alone is sufficient to rule out some DM explanations for the CR excesses, when the excesses are attributed to DM annihilation in the smooth halo.  Any effects that  relax these constraints are of consequence for the viability of these models.  If local substructure is significant, the inner-galaxy constraints are weakened to the point of irrelevance, as we will discuss in Sec.\ref{ssec:innerGalaxy} (see also Sec.\ref{ssec:parametrizing}).

\section{The Effect of Substructure}
\label{sec:effectsofsub}
Having reviewed the constraints on the scenario where the CR excesses originate from Sommerfeld-enhanced DM annihilation in the smooth halo, we now study how the preferred regions of parameter space shift in the presence of local substructure. In all cases, a sufficiently large relic abundance can be readily achieved in these models in conjunction with a fit to the CR excesses, although this constrains the value of $\alpha_D$ when other parameters are fixed.  On the other hand, the CMB and self-interaction constraints can be quite restrictive. While both are independent of the presence of substructure, they are nonetheless sensitive to the same low-velocity physics that is relevant for Sommerfeld enhancement in subhalos. We must be certain that invoking a large low-velocity boost for DM annihilation within subhalos, sufficient to generate an important contribution to the observed positron signal, does not place us in conflict with these observations. While previous studies have considered simply adding substructure to generate a larger signal, this is the first attempt to study the signal self-consistently with robust constraint channels. 

We demonstrate that CMB constraints on the saturated enhancement require that substructure contribute a 40\% enhancement to the local DM density-squared from the smooth halo, in the case where annihilation in substructure provides a good fit to the CR excesses; however, if this condition is satisfied, mediator masses in the $1-200$ MeV range are permitted (in contrast to the case where the local annihilation signal is dominated by the smooth halo). For mediator masses less than $\sim 10$ MeV, the resulting DM-DM scattering cross section exceeds the ``self-interaction threshold'' to significantly affect the structure of dwarf galaxies, but is not ruled out by robust bounds on DM self-interaction.  
If substructure enhances the DM density-squared integral by $\mathcal{O}(1)$, then 
combined local boost factors (from substructure and Sommerfeld enhancement) of $100-1000$ can arise consistently for a wide range of mediator masses, and even in  regions of parameter space where the maximum allowed boost factor from Sommerfeld enhancement in the smooth halo is only $\mathcal{O}(1)$.

\subsection{Parameterizing local substructure}\label{ssec:parametrizing}
The effect of local substructure on a wide variety of constraints can be understood through a simple parametrization, treating the amount of local substructure as a free parameter.  
We write $1 + \Delta(r) = \langle \rho^2(r) \rangle / \langle \rho(r) \rangle^2$, where $r=$ Galactocentric radius, and defer discussion of the expected value of $\Delta$ to the next section.  When  substructure boosts are non-negligible, most of the substructure signal comes from small dense subhalos with saturated Sommerfeld enhancements.
The enhancement factor from substructure and Sommerfeld enhancement can then be written as 
\begin{equation} 
\mathrm{S}_\mathrm{eff}  \approx  S_{v(r)} + S_{v\rightarrow 0} \Delta(r), 
\end{equation}
where $S_v$ is the Sommerfeld enhancement factor at velocity $v$. Generally, one or the other term will dominate, in which case we are either ``smooth halo'' dominated (former term), or substructure dominated (latter term). For cosmic-ray signals, propagation of the CRs from the point of annihilation means that strictly the relevant substructure boost is given by $\Delta(r)$ averaged over some volume. However, for the energies relevant for the PAMELA and \emph{Fermi} signals, most of the observed positrons come from within 1 kpc \cite{Delahaye:2008ua}, and it is reasonable to approximate this average by the local value $\Delta(8.5 \mathrm{kpc})$.

The term ``boost factor'' is commonly used to describe any number of enhancements to the cross section, but generally refers to either the boost from substructure compared to a smooth halo, or the boost by the Sommerfeld enhancement relative to an uncorrected $s$-wave cross section. We will attempt to clearly distinguish between the various ``boosts''
 (in particular because both of these contributions will vary from place to place). We define the general ``boost factor'' (BF) as the enhanced annihilation rate $\langle \sigma v \rho^2  \rangle$ divided by the canonical ($3 \times 10^{-26}$ cm$^3$/s) $\times \langle \rho \rangle^2$, from all combined effects. Then we obtain the relation,
\begin{equation} \mathrm{BF}  =  \mathrm{BF}_\mathrm{smooth} \left(1 + \frac{S_{v\rightarrow 0}}{S_{v(r)}} \Delta(r) \right) = \frac{\langle \sigma v \rangle_{v \sim 150 \mathrm{km/s}}}{3 \times 10^{-26} \mathrm{cm}^3/\mathrm{s}} \left(1 + \frac{S_{v\rightarrow 0}}{S_{v(r)}} \Delta(r) \right) . \end{equation}
If the second term (proportional to $\Delta(r)$) is dominant locally, then one needs to understand its scaling with $r$ -- not just that of the smooth $\rho(r)^2$ -- to understand how limits are affected. 

Consider constraints from the inner Galaxy (or another system with little substructure where the characteristic velocity can be quite high); let us assume that $\Delta(r)=0$ there, i.e. all the substructure has been disrupted. Then the ratio,
\begin{equation} \frac{\mathrm{BF}_\mathrm{GC}}{\mathrm{BF}_\mathrm{local}}  = \frac{\mathrm{BF}_\mathrm{GC,smooth}}{\mathrm{BF}_\mathrm{local,smooth}} \left(1 + \frac{S_{v\rightarrow 0}}{S_{v \sim 150 \mathrm{km/s}}} \Delta(8.5 \mathrm{kpc}) \right)^{-1} = \frac{S_{v(r=0)}}{S_{v \sim 150 \mathrm{km/s}} + S_{v\rightarrow 0} \Delta(8.5 \mathrm{kpc})}
\rightarrow \frac{1}{\Delta(8.5\mathrm{kpc})}\frac{S_{v(r=0)}}{ S_{v\rightarrow 0}}\label{galCompare}
, \end{equation}
is in general not equal to one as most studies of the inner Galaxy limits have assumed.  In the substructure-dominated case, the final expression of \eqref{galCompare} approximates the ratio of boosts, which can easily weaken constraints from the inner Galaxy by up to three orders of magnitude (see Fig. \ref{fig:satlocalratio}) even for moderate $\Delta(8.5 \mathrm{kpc}) \sim 0.1 -1$.

Constraints from systems where the Sommerfeld enhancement is already saturated must also be modified to account for local substructure, and behave quite differently depending on whether the smooth halo or substructure dominates the local signal.  Specifically,
\begin{equation} 
\frac{{\rm BF}_\mathrm{sat}}{{\rm BF}_\mathrm{local}} = \frac{S_{v\rightarrow 0}}{S_{v \sim 150 \mathrm{km/s}} + S_{v\rightarrow 0} \Delta(8.5\mathrm{kpc})}
\rightarrow 
\begin{cases}
{S_{v\rightarrow 0}}/{S_{v \sim 150 \mathrm{km/s}}} & \mbox{smooth-halo-dominated}\\
 1/\Delta(8.5\mathrm{kpc}) & \mbox{substructure-dominated}\\
\end{cases}
\end{equation}
The ratio applicable when the smooth halo dominates can be very large, particularly for mediator masses below 100 MeV, making bounds from the CMB particularly constraining of these models.  In the substructure-dominated case, by contrast, Sommerfeld-enhanced models behave like models with a large but velocity-independent annihilation cross section, with a local density-squared rescaled by $\Delta$.  The resulting constraints are potentially orders of magnitude weaker than one would have thought by considering only the smooth component of the local halo.  

\subsection{Consistent scenarios for thermal freeze-out}
\label{subsec:relicdens}
In general, in the presence of a dominant substructure contribution, the value of $\alpha_D$ that produces the CR excesses is too \emph{small} to generate the observed relic density by thermal freezeout, at least in the simplest models (in contrast to the smooth-halo case with larger DM mass and different decay modes studied in \cite{Feng:2010zp}). The benchmarks given in \cite{Finkbeiner:2010sm}, which achieve the correct relic density and local boost factor assuming the entire signal originates from the smooth halo, are generally not appropriate for the substructure-dominated case because they overproduce the CR signal.  

Instead, one is led to consider models where additional annihilation channels are important during freeze-out, which may or may not be relevant for the CR excesses today.
Inclusion of such processes significantly affects which models are allowed, by breaking the linkage between the Sommerfeld enhancement and the annihilation cross section, and hence allowing extra depletion of the relic density while still producing the desired CR signal. To illustrate the interplay between CMB/self-interaction constraints and present substructure enhancements, we consider two limiting cases: a case with only new ``irrelevant'' annihilations, and a case with only new ``relevant'' annihilations, in the sense of being (ir)relevant to indirect detection.

The first possibility is that there may be extra annihilation channels that are important for freezeout, but irrelevant to the CR excesses today. 
These processes might include $p$-wave suppressed annihilation, channels that experience a repulsive Sommerfeld effect, involvement of excited states that are present in the early universe but have decayed by the present day, very soft annihilation channels (which contribute to \epm signals where the backgrounds are large), or annihilations into invisible channels, such as neutrinos or dark-neutralinos.
 
In this case (new ``irrelevant'' channels), we assume that  the annihilation rate relevant to indirect detection in the Galactic halo is calculable from the parameters $\alpha_D$, $m_\phi$, and $m_\chi$ alone, and take it to be given by the minimal $t$-channel annihilation cross section into dark gauge bosons, $\langle \sigma v \rangle = \pi \alpha_D^2 / m_\chi^2 \times$ the Sommerfeld enhancement. We further assume that the Sommerfeld enhancement is controlled by the same force carriers into which the DM annihilates.

Where the calculated annihilation rate is smaller than required to generate the correct relic density, we assume the difference is made up by these extra annihilation channels which -- for whatever reason -- do not contribute signal to present-day indirect detection experiments. Where this rate is too large to generate the correct relic density, we say that this point is ruled out by the relic density constraint. In other words, we take the relic density to provide an upper bound on the annihilation cross section, rather than fixing its value.

Alternatively, there could be additional annihilation channels that contribute to both freezeout and the local CR signal, but which do not correspond to force carriers mediating additional Sommerfeld enhancement.
This could arise for instance by annihilations into an additional force carrier that is not as effective for Sommerfeld enhancement (e.g. due to a more massive force carrier), but has a larger coupling. 
In this class of scenarios (new ``relevant'' channels), we again take $\alpha_D$, $m_\phi$, and $m_\chi$, and assume that these parameters determine the Sommerfeld enhancement, but that the underlying cross section relevant for both freezeout and indirect detection experiments $\langle \sigma v \rangle$, which is Sommerfeld-enhanced at low velocities, can be larger than the naive $ \pi \alpha_D^2 / m_\chi^2$. When even the ``bare'' rate of $\langle \sigma v \rangle = \pi \alpha_D^2 / m_\chi^2$ alone is larger than allowed by the relic density, we again say that this point is ruled out by the relic density constraint (again, taking the relic density to give an upper bound on the cross section).

In both cases, we must take into account the fact that the presence of Sommerfeld enhancement reduces the ``bare'' annihilation cross section that gives the correct relic density \cite{Dent:2009bv, Zavala:2009mi, Feng:2010zp, Finkbeiner:2010sm}. We follow \cite{Essig:2010em} and use a Taylor expansion of the Sommerfeld enhancement during freezeout to estimate the magnitude of this effect.

\subsection{Self-consistent substructure-dominated scenarios}

\begin{figure*}
\includegraphics[width=0.45\textwidth]{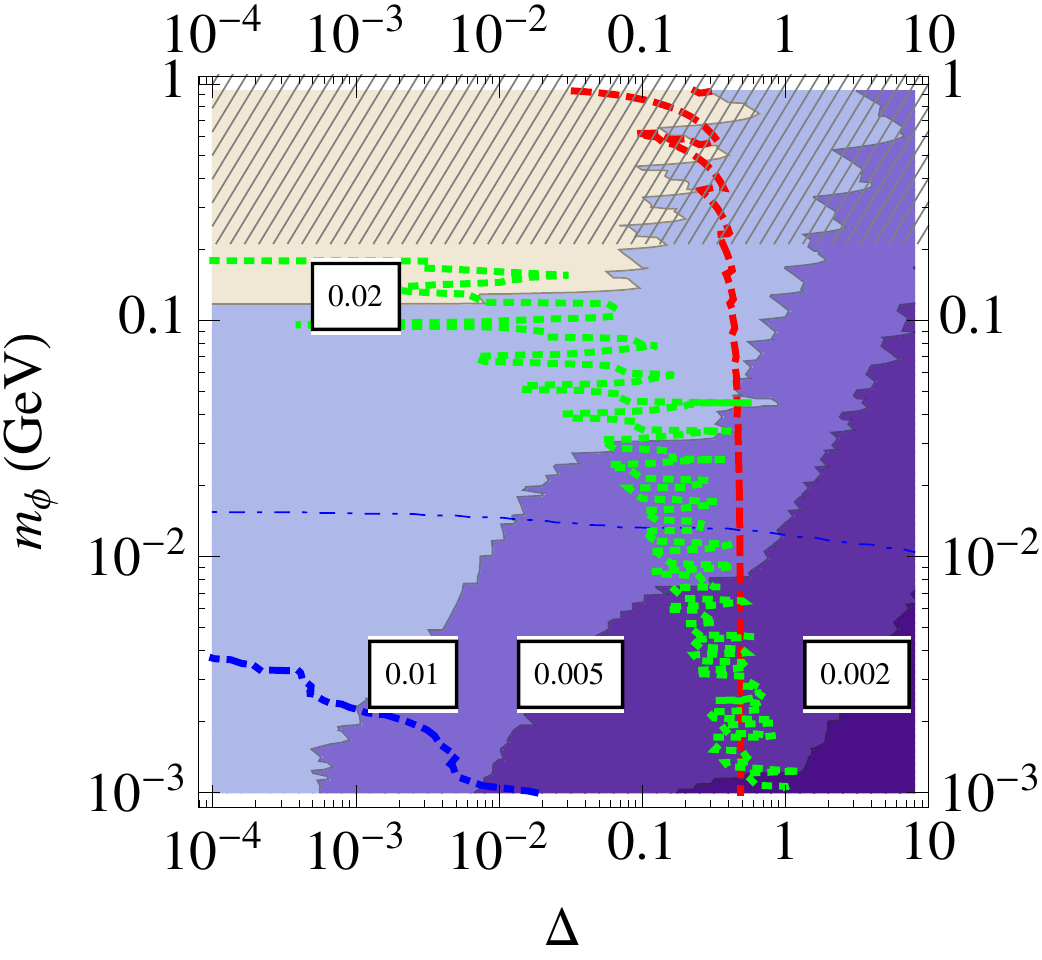}
\includegraphics[width=0.45\textwidth]{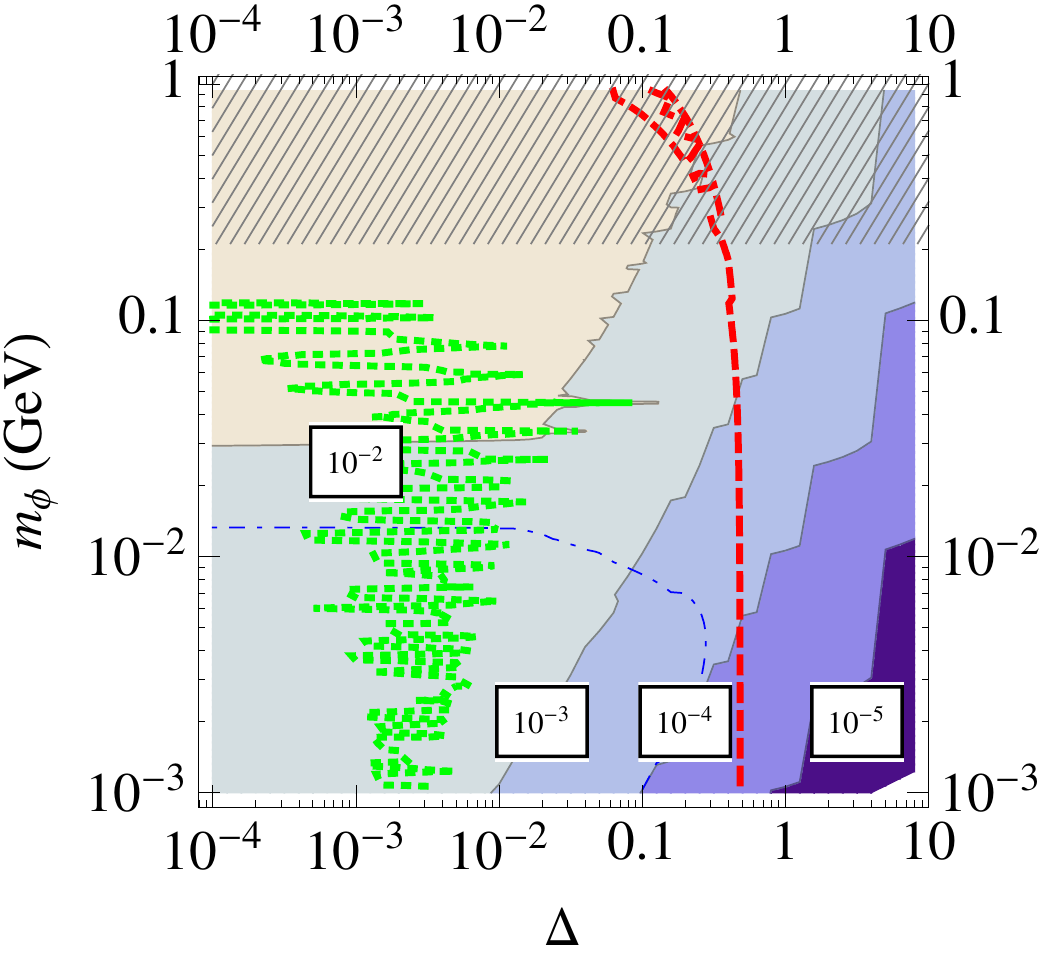}
\caption{Contours of constant dark sector coupling $\alpha_D$ as a function of mediator mass $m_\phi$ and substructure contribution $\Delta$, for a fixed DM mass of 1.2 TeV and local boost factor (BF) of 100, in the scenarios with new ``irrelevant channels'' (\emph{left panel}) and ``relevant channels'' (\emph{right panel}). The BF includes contributions from Sommerfeld enhancement and substructure and is defined by $\mathrm{BF}  = \frac{\langle \sigma v \rangle_{v \sim 150 \mathrm{km/s}}}{3 \times 10^{-26} \mathrm{cm}^3/\mathrm{s}} \left(1 + \frac{S(v\rightarrow 0)}{S(v \sim 150 \mathrm{km/s})} \Delta(r\sim 8.5\mathrm{kpc}) \right)$.  Regions to the left of and/or below the red dashed (blue dot-dashed) lines are ruled out by constraints from the CMB (self-interaction bounds).  The thin blue dot-dashed line denotes the threshold at which self-interaction effects in dwarf galaxies become significant, but may be allowed.
The region to the left and below the green dotted line is where WIMPonium formation becomes relevant; at this DM mass and for this class of models, WIMPonium formation is almost always ruled out by the CMB constraints, although it can become relevant at higher DM masses. The dark gauge boson is assumed to decay into electrons only, in which case this boost factor and DM mass provide a good fit to the PAMELA and \emph{Fermi} data. When the gauge boson mass exceeds twice the muon mass, the true final state may become more complicated, so this region is indicated by cross-hatching. For all points on this plot, the Boltzmann equation has been solved numerically to confirm that the thermal relic density is not over-depleted.}
\label{fig:compareconstraints}
\end{figure*}

We can now choose the parameters to achieve a desired boost factor locally -- in particular, one consistent with the CR excesses -- and ask how constraints from the CMB and self-interaction limits bound the force carrier mass, as a function of the substructure contribution. As previously, we consider the Sommerfeld enhancement induced by a Yukawa potential (a slightly more complex example is discussed briefly in Appendix \ref{app:inelastic}), as this simple case adequately demonstrates the effect of non-zero $\Delta$. We use a DM mass of 1.2 TeV and a local boost factor of 100 as a benchmark, and assume the dark force carrier decays only into electrons, as appropriate for a vector boson with mass $< 2 m_\mu$: this benchmark provides an adequate fit to the PAMELA and \emph{Fermi} measurements. It is sometimes claimed that the electron-only channel gives a sharp spectral endpoint inconsistent with the CR data, but while this is true for \emph{direct} DM annihilation to $e^+ e^-$, annihilation to $\phi \phi$ followed by $\phi \rightarrow e^+ e^-$ decay gives an endpoint of similar sharpness to $\chi \chi \rightarrow \mu^+ \mu^-$, and provides a good fit to the data \cite{Meade:2009iu} (an example model with slightly lower DM mass and boost factor is shown in \cite{Finkbeiner:2010sm}). Showering or decays within a more complex dark sector can also soften the endpoint and improve the fit to the data \cite{Meade:2009iu, Mardon:2009rc}.

For each point in $\Delta-m_\phi$ parameter space, we numerically solve for the largest value of the dark sector coupling $\alpha_D$ that gives rise to the desired local boost factor:  smaller values of $\alpha_D$ may achieve the same local boost factor, but only if they lie near the tips of narrow resonance peaks (a requirement which becomes increasingly finely tuned as $\alpha_D$ decreases, since the enhancement must be increasingly close to perfectly resonant to cancel out the usual reduction in the enhancement from lowered $\alpha_D$). Consequently, choosing the largest possible $\alpha_D$ yields the most ``natural'' parameter set giving the desired boost factor, in the sense that small changes to the parameters will not change the boost factor very much. The resulting self-interaction cross section and signal in the CMB can then be compared to the limits. There is a third constraint from relic density, as described above for the cases of extra relevant and irrelevant channels, which we check by solving the Boltzmann equation for the late-time DM density (including Sommerfeld enhancement). However, for a required boost factor of 100 at $m_\chi = 1.2$ TeV, the constraint curve does not appear on these plots. Even for zero local substructure and zero mediator mass, a boost factor of 100 is attainable via the Coulomb-like $\pi \alpha_D/v$ Sommerfeld enhancement while maintaining consistency with the relic density bound (as can be seen from e.g. \cite{Feng:2010zp}). Our results are shown in Fig. \ref{fig:compareconstraints}.

The major difference between the two scenarios we consider (with extra ``irrelevant'' and ``relevant'' channels respectively) is the rate at which the preferred value of $\alpha_D$ falls with increasing $\Delta$, and hence the strength of the self-interaction bound at low masses. In the first case, with extra irrelevant annihilation channels, the saturated annihilation rate scales as $\alpha_D^3$, whereas in the case with additional relevant channels, the bare annihilation rate is largely fixed by the relic density alone, and so the saturated annihilation rate scales roughly as $\alpha_D$. Consequently, in the ``irrelevant channels'' case relatively small changes to $\alpha_D$ are sufficient to greatly reduce the signal, compensating for the increased boost factor from saturated enhancement in subhalos; $\alpha_D$ changes by only a single order of magnitude over the parameter space we consider. If the self-interaction threshold discussed earlier is treated as a limit, it remains quite stringent at mediator masses below $m_\phi \sim 10$ MeV. In the ``relevant channels'' scenario, on the other hand, a large reduction in the saturated annihilation rate requires a large reduction in $\alpha_D$: the very low values of $\alpha_D$ at low mediator mass and $\mathcal{O}(1)$ $\Delta$ also greatly relax any self-interaction bounds.

If the force carrier mass is sufficiently light, $m_\phi < \alpha_D^2 m_\chi/4$, then it is possible for two DM particles to radiatively capture into a bound state at low velocities, referred to as WIMPonium. The capture cross section scales in the same way as Sommerfeld-enhanced annihilation in the low-velocity limit, but is larger by a factor of $\sim 6$ in the limit where $m_\phi \ll  \alpha_D^2 m_\chi/4$. For this DM mass, WIMPonium formation primarily affects regions of parameter space that are already ruled out by the CMB bounds, but it can be marginally relevant for $\Delta \sim 0.5-1$ and few-MeV force carriers, and is included in the plots.

Fig. \ref{fig:compareconstraints} is useful for showing how the different constraints compare, but relies on picking a specific target boost factor. A question of perhaps more general interest is how the maximal boost factor \emph{consistent with all constraints} varies as a function of $\Delta$ and $m_\phi$. We again proceed by sampling the $\Delta-m_\phi$ parameter space holding $m_\chi$ fixed at 1.2 TeV, and at each point scan over $\alpha_D$ and numerically check consistency with the CMB, self-interaction and relic density bounds, to obtain the maximum boost factor consistent with these limits. We include radiative capture to WIMPonium in the boost factor, for values of $\alpha_D$ where it is kinematically allowed. 

As mentioned previously, at resonance peaks relatively low values of $\alpha_D$ can give rise to very large saturated enhancements. Relying on such resonance peaks is problematic for several reasons:
\begin{itemize}
\item The usual treatment of the Sommerfeld enhancement neglects higher-order corrections that regulate the resonances; the saturated enhancement in our current approximate treatment diverges at the exact centers of the resonances, and this is not physical.
\item On the resonances, the enhancement saturates at lower velocities than in the non-resonant case; close to the centers of the resonances, it is not clear that we can assume the enhancement is saturated in the smallest subhalos.
\item Our perturbative treatment of Sommerfeld corrections to thermal freezeout fails in the case of highly resonant enhancement, since in this case annihilations can recouple after kinetic decoupling.
\item From an aesthetic perspective, demanding that the Sommerfeld enhancement be highly resonant implies fine-tuning of the parameters.
\end{itemize}
Consequently, we impose the further condition that the saturated Sommerfeld enhancement must not exceed the expected non-resonant value, $12 \alpha_D m_\chi / m_\phi$, by more than a factor of 10. Increasing this factor to 100 (or decreasing it to 3) has a negligible effect on the results.

\begin{figure*}[ht]
\includegraphics[width=0.45\textwidth]{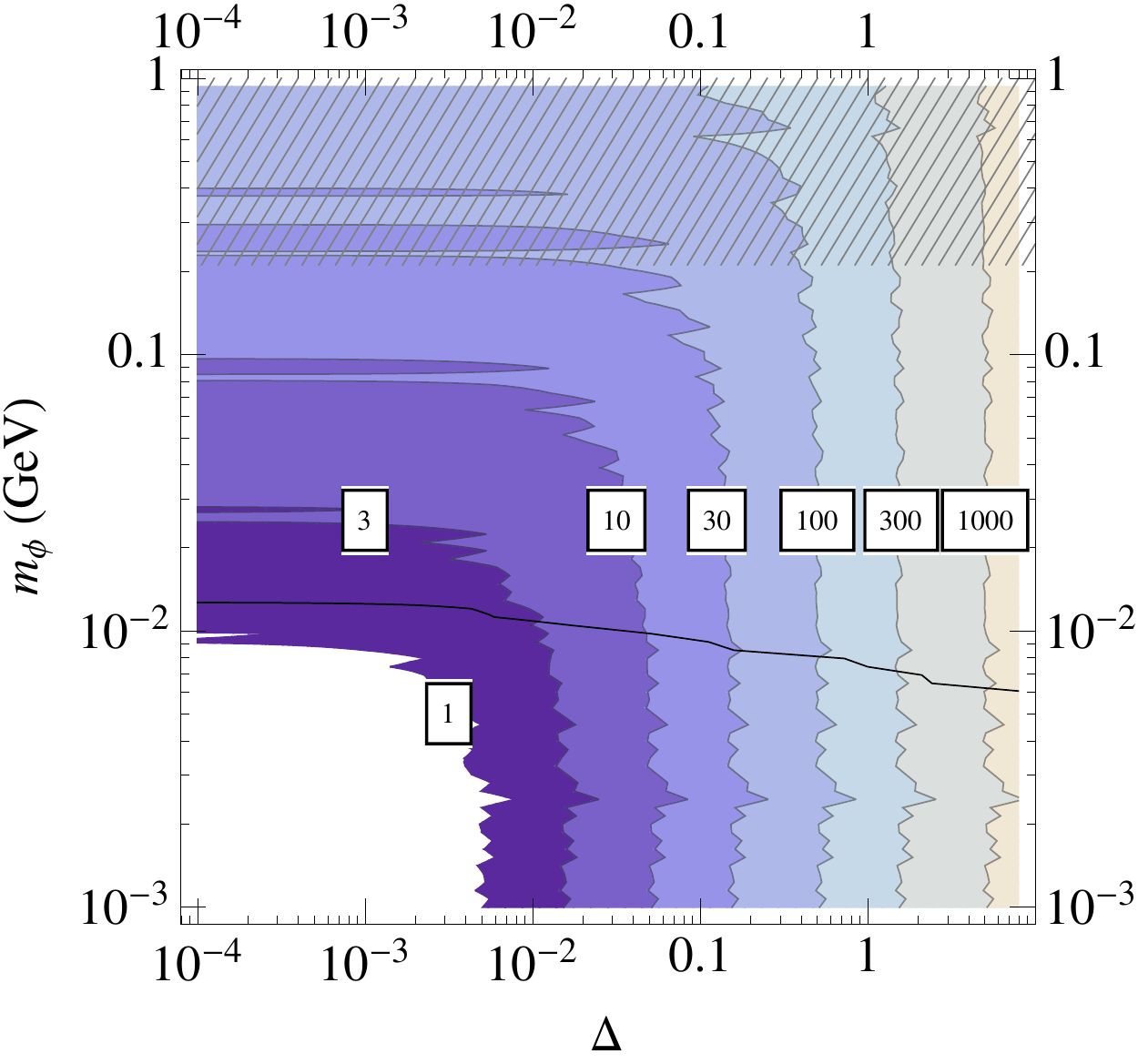}
\includegraphics[width=0.45\textwidth]{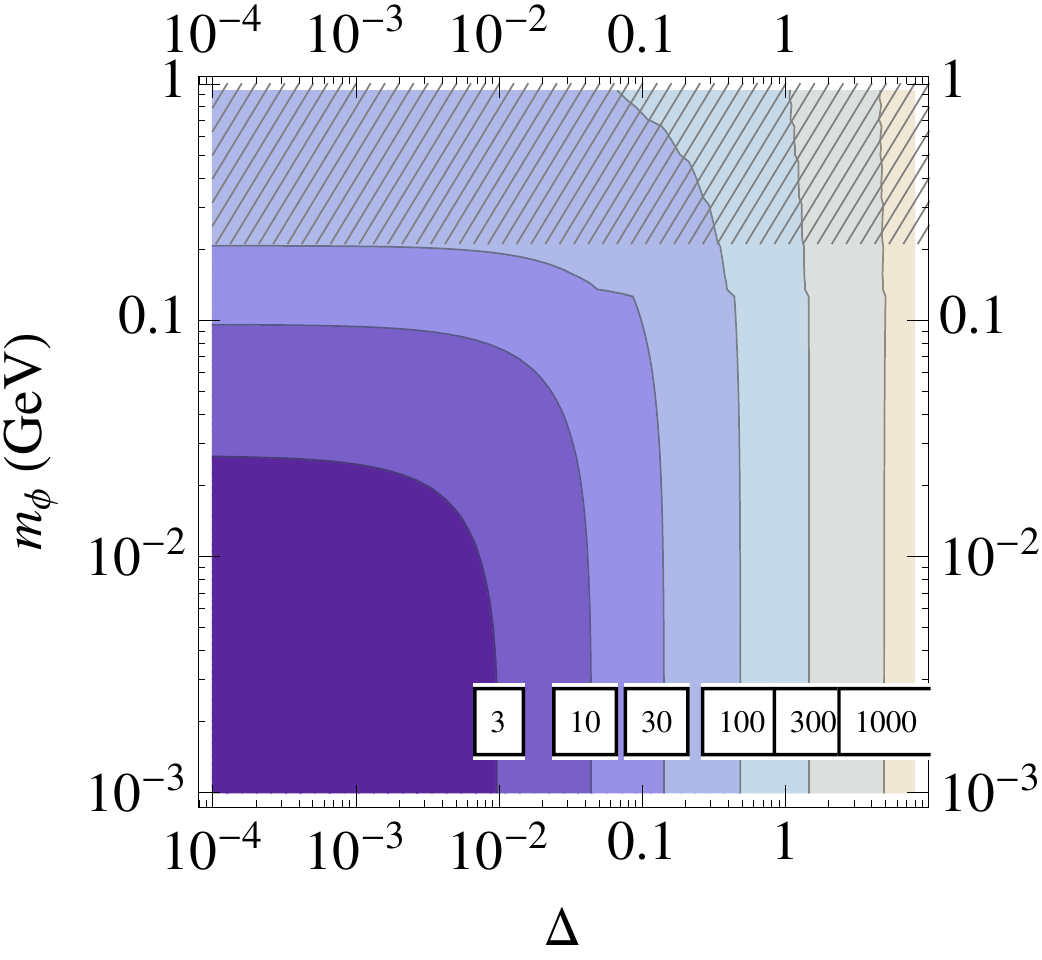}
\caption{The maximum local boost factor for 1.2 TeV DM consistent with constraints from the thermal relic density, the CMB, self-interaction bounds, and naturalness (in the sense of not relying on the resonance peaks), in the scenarios with new ``irrelevant channels'' (\emph{left panel}) and ``relevant channels'' (\emph{right panel}). The dark gauge boson is assumed to decay into electrons only; when the gauge boson mass exceeds twice the muon mass, the true final state may become more complicated, so this region is indicated by cross-hatching.  In the left panel, parameter points that maximize the boost for $m_\phi < 13$ MeV have transfer cross-sections at $10 \km/\s$ above the ``self-interaction threshold'' of $0.1 \cm^2/\g$.  Treating this threshold as a hard constraint would extend the white region ($\rm{BF}_{\rm local,max} \le 1$) out to the black curve, while having no effect at all on the contours for $m_\phi> 13\MeV$.  In contrast, the self-interaction bounds are never constraining for scenarios with new relevant channels.}
\label{fig:maxlocalboost}
\end{figure*}

The results of this analysis for scenarios with either ``irrelevant'' or ``relevant'' annihilation channels are shown in Fig. \ref{fig:maxlocalboost}.  The limiting constraint in most of the parameter space is that from the CMB, with dwarf evaporation becoming a significant constraint only at low $\Delta$ and $m_\phi$, in the case of ``irrelevant channels'' (which require larger $\alpha_D$ to produce a given boost).   The more stringent ``self-interaction threshold'' of $\sigma_T = 0.1 \cm^2/\g$ at $v_0 = 10$ km/s, above which DM scattering within the dwarf  halo would change its shape significantly from non-interacting CDM simulations, is not included as a constraint, but the parameter points shown in the left plot (irrelevant channels) cross this threshold at $m_\phi = 13$ MeV.  If we were to treat this threshold as a constraint, it would leave the region $m_\phi>13 \MeV$ completely unaffected, but sharply change the maximum boosts in the region $m_\phi < 13 \MeV$, with the white region $BF_{\rm local}\le1$ pushed out to the near-horizontal black line.  No self-interaction effects are important in the ``relevant channels'' scenario because of the much slower scaling of the annihilation rate with $\alpha_D$. 

In Fig. \ref{fig:maxlocalboost} we see that a maximal boost factor of $\sim 100$ is first achieved at $\Delta \sim 0.4-0.5$ for small mediator masses; in this region of parameter space, the CMB provides the strongest constraint. This is consistent with Fig. \ref{fig:compareconstraints}, where we see that the target boost factor of 100 is first consistent with the CMB limits at $\Delta \sim 0.4-0.5$. We have checked the effect of running the analysis with and without including WIMPonium formation and found that it makes essentially no difference to our results (although the plots we show do include it): the parameters for which WIMPonium formation is kinematically allowed are in general excluded by the constraints, or at most marginally allowed, and so contribute little to the maximum boost factor.

\begin{figure*}[ht]
\includegraphics[width=0.45\textwidth]{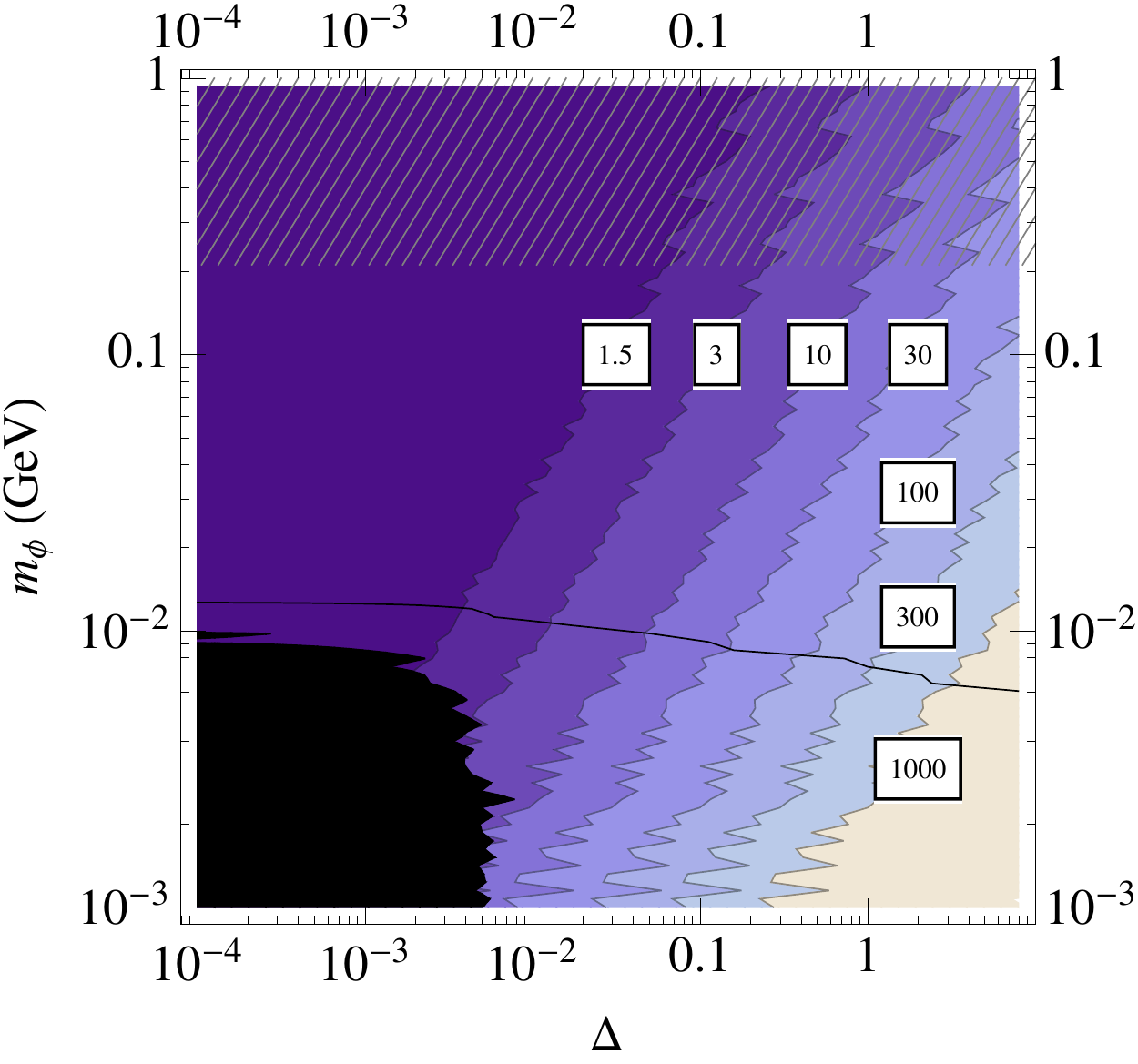}
\includegraphics[width=0.45\textwidth]{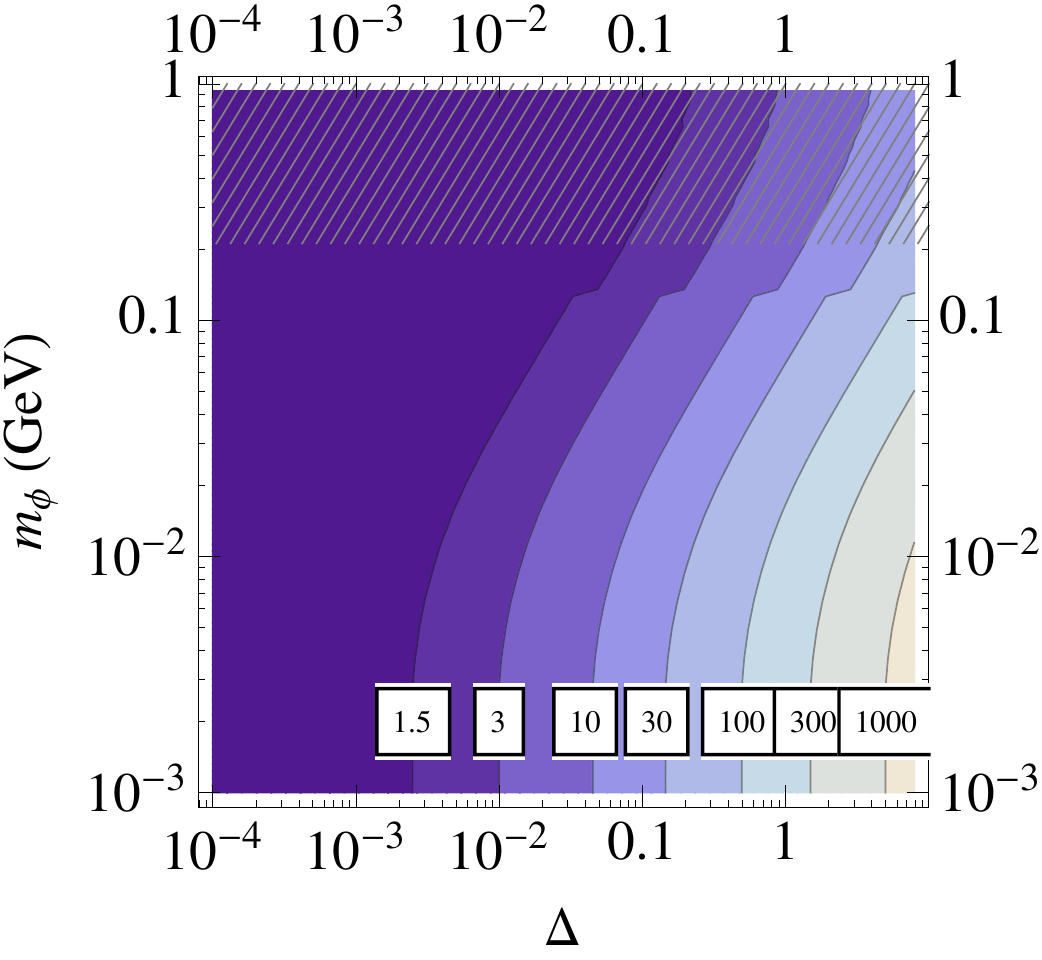}
\caption{The ratio of the maximum local boost factor for 1.2 TeV DM to the maximum local boost factor with $\Delta = 0$, where in both cases the parameters achieving the maximum local boost are required to respect constraints from the thermal relic density, the CMB, self-interaction bounds, and naturalness (in the sense of not relying on the resonance peaks), in the ``irrelevant channels'' (\emph{left panel}) and ``relevant channels'' (\emph{right panel}) scenarios. The dark gauge boson is assumed to decay into electrons only; when the gauge boson mass exceeds twice the muon mass, the true final state may become more complicated, so this region is indicated by cross-hatching. In the ``irrelevant channels'' scenario, the region where the substructure-enhanced local boost is less than 1 is blacked out, since while this region may have a larger boost factor with substructure than without, the boost factor is still very small and so its details are not very interesting.
If the ``self-interaction threshold'' from self-interaction of DM within dwarf galaxies were treated as a hard constraint, this  black region would extend out to the black curve.
}
\label{fig:relativeboost}
\end{figure*}

In the $\Delta=0$ limit, the maximum boost factor is strongly dependent on $m_\phi$, but for $\Delta$  of $\mathcal{O}(1)$ this dependence is almost completely removed.  
This is to be expected: in this regime the CMB provides the strongest limits, the local enhancement is substructure-dominated, and the ratio of the local enhancement to the saturated enhancement depends only on $\Delta$. 
(We remind the reader, however, that models saturating the CMB limit with very low $m_\phi \lesssim 10$ MeV have transfer cross-sections of order $0.1 -1 \cm^2/\g$ at 10 km/s, whose effect on the substructure of dwarf galaxies is significant.)
Fig. \ref{fig:relativeboost} shows the increase in maximum boost factor (consistent with all constraints) in the presence of substructure, as a function of $\Delta$ and $m_\phi$: even for  $\Delta$ of $\mathcal{O}(1)$, the factor can be two or more orders of magnitude.

\section{The Distribution of Substructure and its Implications}
\label{sec:constwsub}

Having explored the effect of non-zero local substructure $\Delta$ on the CR signal, we would like to understand both the most likely value of $\Delta$ and the implications of non-zero $\Delta$ for other possible constraint channels. However, the amount of local substructure is highly uncertain. $N$-body simulations cannot resolve the small subhalos that are expected to contribute the bulk of the signal, so some extrapolation procedure must be employed (over 12 or more orders of magnitude). The mass of the smallest subhalos scales as the cube of the temperature of kinetic decoupling of the DM from the SM, which can easily range from $1-100$ MeV. Furthermore, DM self-interactions and baryonic physics, neither of which are included in $N$-body simulations, may deplete substructure and/or flatten the density profiles of subhalos at later times. 

Consequently, it is important to explore the consequences of a broad range of substructure scenarios. We consider the formulations of \cite{Kamionkowski:2010mi,Pieri:2009je,Kistler:2009xf}, each of which attempt to understand the implications of $N$-body simulations, and study their consequences for a range of gamma-ray signals. These signals have been considered previously in other papers, but not generally in the context of a scenario where the local signal is substructure-dominated due to Sommerfeld enhancement.

We examine the effects of abundant substructure on constraints from the inner MW, dwarf galaxies, and the extragalactic gamma-ray background. We find that the inner-Galaxy constraints can become nearly irrelevant in substructure-dominated scenarios.  While constraints from dwarfs are likely strengthened by substructure (relative to the case that the local and dwarf halos are both dominated by their smooth components), the parameter space of interest remains allowed. Bounds from the extragalactic gamma-ray background radiation are more stringent, requiring a local substructure enhancement $\Delta \gtrsim 1$ in the substructure-dominated scenario.  These bounds are most easily accommodated with $\Delta \sim 6$, but smaller $\Delta$ can be consistent within reasonable uncertainties in the abundance of isolated small halos and the DM damping scale.

\subsection{Unresolved substructure near and far}
There are many different approaches to the question of substructure boosts, often relating to whether the study focuses on the local boost, the boost to emission from a distant source (such as a dwarf galaxy), or the boost to the extragalactic, all-halo and all-redshift, isotropic signal. However, in general, there are three parameters which influence the relevant substructure boost: $1-f_{sm}$, the fraction of the DM in bound substructures ($f_{sm}$ is the fraction in the ``smooth'' halo), $\alpha_{sub}$, the exponent of the power law that controls the distribution of subhalos of different masses\footnote{In different formulations, this can describe the distribution of subhalos in main halos \cite{Zavala:2011tt,Pieri:2009je} or the probability of finding a particle in a region of density $\rho$ (related to the number of small subhalos of that characteristic density) \cite{Kamionkowski:2010mi}.}, and finally $\alpha_h$, which is the parameter which determines the distribution of unresolved {\em main} halos, relevant for studies of isotropic diffuse photons.

The local boost is most dependent on a) the total amount of substructure as well as the mass distribution of subhalos controlled by $\alpha_{sub}$, and b) the amount of substructure present in the solar neighborhood $1-f_{sm}$. The boost in the outer part of the MW is controlled by essentially the same two parameters. Thus, the ratio between the two is dominantly controlled by the evolution of  $1-f_{sm}$ as a function of radius (in addition to the change in concentration as a function of radius from tidal stripping \cite{Pieri:2009je}). 

\begin{figure*}
\includegraphics[width=0.45\textwidth]{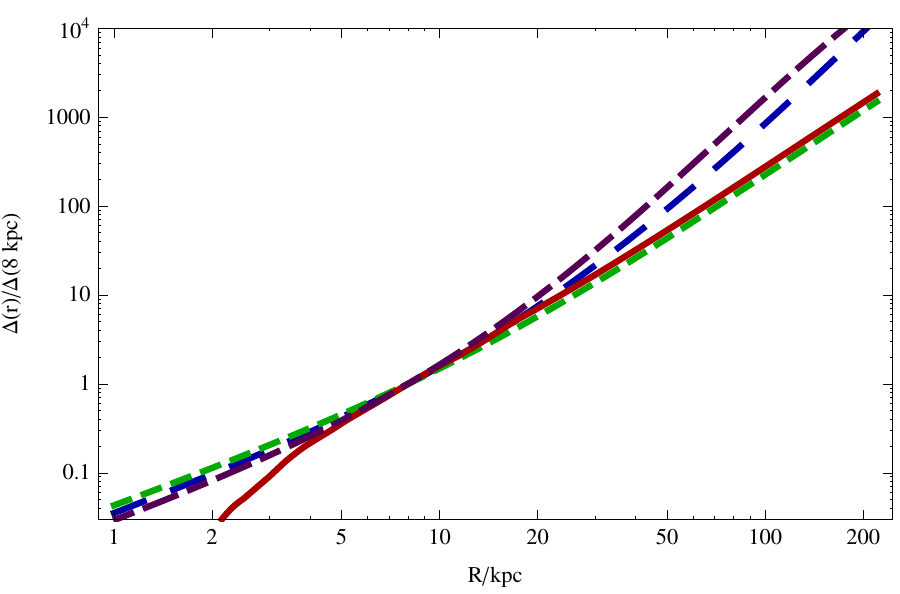}
\includegraphics[width=0.45\textwidth]{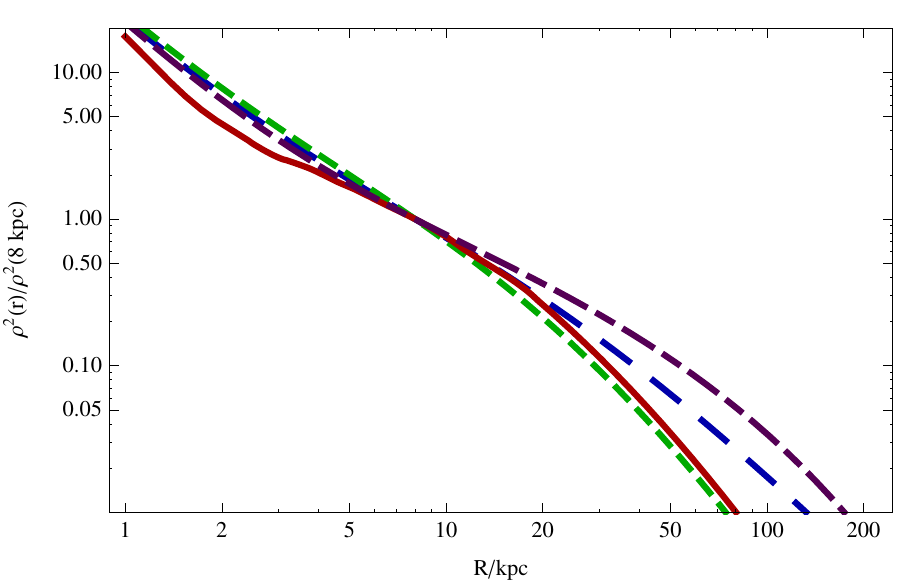}
\caption{{\it left}: the profile  $\Delta(r)/\Delta(8.5 \mathrm{kpc})$ for four different approaches to extrapolating $N$-body results into the inner halo. {\it right}: $\rho^2(r)/\rho^2(8.5 \mathrm{kpc})$ (relevant for line-of-sight calculations) with $\Delta(8.5 \mathrm{kpc}) \times \sigma v_{sat}/\sigma v_{smooth} = 10$ for the same extrapolations. Lines are: the approach of \cite{Kamionkowski:2010mi} to the Via Lactea II simulation ({\it \color{blue} blue, long-dashed}), the approach of \cite{Kistler:2009xf} to the Aquarius simulations ({\it \color[rgb]{0.7,0,0.7} purple, long-short dashed}), the approach of \cite{Pieri:2009je} to the VLII simulation, with tidal disruption ({\it \color{red} red, solid}), and without tidal disruption ({\it \color{darkgreen} green, dashed}).}
\label{fig:innerouterhalo}
\end{figure*}

\subsection{Inner Galaxy gamma-ray flux}\label{ssec:innerGalaxy}

While various approaches differ on the rate at which the substructure signal is disrupted towards the GC, there is general agreement that it {\em should} be suppressed as one moves to smaller radii.  Thus, in the limit that the local signal is dominated by substructure, one can estimate the maximum amount by which any inner galaxy signals would be suppressed (compared to the no substructure case) by looking at the evolution of $\Delta(r)$, which we show in Fig. \ref{fig:innerouterhalo}. Even in the case with the slowest evolution of substructure of the four cases we consider (from \cite{Pieri:2009je} without tidal disruption), the signal is suppressed in the inner 1 kpc by a factor of $\sim$20 (although of course, there will be some contribution from the smooth component as well, which does not suffer this suppression). Thus, in the limit that local substructure dominates the PAMELA signal {\em even the most stringent ICS or FSR constraints can become irrelevant.}

\subsection{Outer halo gamma-ray flux}

At the same time, one might be concerned that by boosting the local substructure signal, one is also boosting {\em other} signals that are already dominated by substructure, namely, gamma rays from the outer galaxy and unresolved extragalactic sources (the latter contributing to the isotropic diffuse gamma-ray flux). The limits from gamma-ray emission from dwarf galaxies, briefly mentioned previously, will also become more stringent if substructure is taken into account.

For the outer galaxy, there are already studies \cite{Kistler:2009xf,Abazajian:2010zb} that consider the role of substructure in ICS signals from the outer halo. \cite{Kistler:2009xf} employs the Aquarius simulation, normalizing the local signal assuming that the entire local PAMELA signal arises from substructure (i.e., $S_{v\rightarrow 0} \Delta \gg S_{v \sim 150 \mathrm{km/s}}$), while \cite{Abazajian:2010zb} uses the approach of \cite{Kamionkowski:2010mi} to the VLII simulation, with a local value $\Delta \sim 0.5$, and implicitly takes $S_{v\rightarrow 0}=S_{v \sim 150 \mathrm{km/s}}$. Both papers calculate the ICS signals from the outer halo and use this to set constraints. 

If we assume that the local signal is substructure dominated, then the outer halo is substructure dominated as well. In this limit, only the product $\Delta(r) S(v\rightarrow 0)$ is relevant. Since \cite{Kistler:2009xf} assumes substructure domination, their limits are directly applicable here. On the other hand, since \cite{Abazajian:2010zb} takes $\Delta(8.5 \kpc) \sim 0.5$, their limits should be strengthened by a factor of $\sim 2$ in the substructure-dominated case. Importantly, both analyses assume that the $e^+e^-$ produce their ICS signals at the point of annihilation. Since the energy loss time for these particles is $\mathcal{O}$(Myr), the particles would at least partially diffuse away, and suppress these limits by up to $\mathcal{O}$(few). 

Even without accounting for this correction, \cite{Kistler:2009xf} finds these signals are only borderline, not excluded. Moreover, observing Fig. \ref{fig:innerouterhalo}, we see that, of the three approaches we have considered, in this approach the substructure is depleted most rapidly as one moves in from the outer halo.   
Other formulations of substructure, with weaker dependence on Galactocentric radius, will yield weaker constraints from the outer MW halo, relative to a fixed local signal.   In light of this uncertainty and the effects of  \epm diffusion, we conclude the the ICS signals of the outer MW halo do not strongly constrain the substructure dominated scenario.

\subsection{Gamma-rays from dwarfs}
\label{subsec:dwarfs}

The non-observation of a gamma ray excess from the dwarf galaxy Segue 1 constrains the DM annihilation cross section to be no more than $\sim 100$ times larger than required to fit the CR excesses with $4e$ annihilation (neglecting substructure both locally and in the dwarf) \cite{Essig:2010em}.  The presence of substructure in both systems would enhance the annihilation signal from Segue 1 more than the local CR signal, because the full substructure boost to Segue 1 is dominated by its substructure-rich outer halo.   However, this ratio is not likely to provide the two-order-of-magnitude relative enhancement to the Segue 1 signal that would be needed to derive a strong constraint from the current measurement.
We may expect the overall boost factor for Segue 1 to be at most comparable to that of a MW-like galaxy, and plausibly smaller because the dwarf possesses substructure over fewer decades of mass.  As discussed below, 
we find that the total boost of the MW seen from far away exceeds the local boost $\Delta$ by a factor of $\sim 5-30$, following the approaches of \cite{Pieri:2009je} and \cite{Kamionkowski:2010mi}.  A comparable boost to the Segue 1 signal would not be sufficient to derive a strong constraint.  

\subsection{Diffuse extragalactic gamma-rays}

A more stringent constraint involving substructure is the limit from isotropic diffuse gamma rays \cite{Abdo:2010dk, Hutsi:2010ai, Zavala:2011tt}. The distribution of main halos, and the distribution of substructure (by mass), are critical for this constraint, but the limit is essentially divorced from the question of how the substructure evolves with Galactocentric radius in the inner galaxy, which is key for local signals. In a recent analysis, \cite{Zavala:2011tt} employed the  approach previously developed by \cite{Zavala:2009zr}, where two independent quantities are modeled as power laws with parameters fitted from $N$-body simulations: the boost from main halos below the resolution limit of the simulation, and the boost associated with each main halo due to its substructure, as a function of the main halo mass. 

For the extragalactic diffuse gamma-ray signal, we take as our comparison point the integrated flux from all smooth main halos with masses between $6.89 \times 10^8 h^{-1} M_\odot$ and $10^{15} M_\odot$. The lower limit corresponds to the resolution of the Millennium-II simulation, and the power-law behavior described in  \cite{Zavala:2009zr}  may no longer be accurate for halo masses $\gtrsim 10^{15} M_\odot$. All boosts for the extragalactic diffuse signal are ``scaled isotropic boosts,'' defined with respect to this quantity.

The scaled boost factor to the extragalactic diffuse signal from the combined substructure and unresolved main halo contributions then ranges from $\sim 20-2500$ when varying the parameters $A_\mathrm{sub}$ and $\alpha_\mathrm{sub}$, which respectively determine the normalization and slope of the power law governing the substructure mass function, across the ranges $10^{-0.5} \le A_\mathrm{sub} \le 10^{0.1}$, $-1.15 < \alpha_\mathrm{sub} < -0.95$. This range of scaled isotropic boosts assumes the parameters of the power law governing the unresolved main halos are held fixed at their best-fit values, with a fixed minimum subhalo mass of $10^{-6}$ $M_\odot$.

In the case with the \emph{smallest} amount of unresolved substructure within the included parameter space, corresponding to a scaled boost to the isotropic signal of $\sim 20$ by our definition, \cite{Zavala:2011tt} have shown that the maximum \emph{saturated} annihilation cross section is characteristically very close to that which is required to explain PAMELA in the absence of substructure, even before any subtraction of astrophysical backgrounds (the uncertainties on the backgrounds are sufficiently large that their contribution to the signal could very well be subdominant). In other words, in substructure-dominated scenarios we would simultaneously require the local boost $\Delta(8.5\mathrm{kpc}) \gtrsim 1$ when the total scaled boost to the extragalactic emission is $\sim 20$. More generally, any scenario where  $\Delta(8.5\mathrm{kpc}) / (\mbox{scaled isotropic boost}) \gtrsim 1/20$ could simultaneously fit PAMELA and evade this constraint. A question we wish to confront is: is such a ratio of substructure boosts reasonable?

It is difficult to address this question directly within the framework of  \cite{Zavala:2011tt}, because the simulations used there cannot resolve structures below $\sim 10^8 M_\odot$, and do not address local boosts deep inside the Galactic halo. On the other hand, the other methods in the literature are not readily expressed in the parameterization of \cite{Zavala:2009zr}. However, we note that for the models we have tested, the total integrated boost from a distant halo is essentially independent of behavior in the inner Galaxy (being dominated entirely by the outer halo); furthermore, if we follow  \cite{Pieri:2009je} and use the Roche criterion to parameterize tidal disruption of substructure, the resulting reduction in the inner-Galaxy boost factor is nearly independent of the parameters $\alpha_\mathrm{sub}$, $A_\mathrm{sub}$. Thus, it makes sense to treat the radial profile of substructure in the inner Galaxy and the overall normalization of the boost factor as unrelated, and simply use the results from \cite{Pieri:2009je, Kamionkowski:2010mi} to obtain the ratio of the local ($R \sim 8.5$ kpc) boost to the overall boost of the halo. Thus, to determine the ratio of the local boost to the isotropic boost and whether it is larger than $1/20$, we take
\be
\frac{\Delta(8.5 \kpc)}{\mathrm{BF}_\mathrm{isotropic}} \simeq \left(\frac{\Delta(8.5 \kpc)}{\mathrm{BF}_\mathrm{MW}} \right)_\mathrm{out\rightarrow in}\left(\frac{\mathrm{BF}_\mathrm{MW}}{\mathrm{BF}_\mathrm{isotropic}} \right)_\mathrm{ZSB}.
\ee
Here the first factor is calculated using the approaches \cite{Pieri:2009je} or \cite{Kamionkowski:2010mi} to connect the local inner boost to the ``total boost'' of the MW as seen from far away (dominated by outer halo structure), while the second factor is calculated using the approach of Zavala, Springel and Boylan-Kolchin \cite{Zavala:2009zr} to connect the MW total boost to the scaled isotropic boost.

Using the approach of  \cite{Pieri:2009je} for the first factor, we find that the total boost of the MW, integrating out to 200 kpc, is approximately $6 \times$ the local value of $\Delta$, 
while for \cite{Kamionkowski:2010mi} the corresponding factor is 18. 
Thus the extragalactic gamma ray bounds can be evaded if the scaled diffuse gamma-ray boost, integrated over all halos (up to $10^{15} M_\odot$) and all substructure, is $\lesssim 1-3 \times$ greater than the total boost of the MW.

Working in the formalism of \cite{Zavala:2009zr}, we can now determine that this condition naturally holds for $\alpha_\mathrm{sub} \lesssim -1.00$ for the largest values of $A_\mathrm{sub}$ and $\alpha_\mathrm{sub} \lesssim -1.05$ for the smallest $A_\mathrm{sub}$. Note that this is \emph{not} the low-substructure limit: as an example, taking the central values $\alpha_\mathrm{sub} = -1.05$, $A_\mathrm{sub} = 10^{-0.2}$, the boost factor for a 10$^{12}$ solar mass halo is $\sim 35$ (corresponding to a local $\Delta \sim 6$ under the formalism of  \cite{Pieri:2009je}, or $\Delta \sim 2$ under the formalism of \cite{Kamionkowski:2010mi}), and for the extragalactic diffuse signal is $\sim 90$. Thus, for these cases, we estimate $\Delta(8.5\kpc)/\mathrm{BF}_\mathrm{isotropic} \sim 1/15$ and $1/50$. 

So far, we have assumed the parameters of the unresolved $\emph{main}$ halos and the cutoff mass are perfectly known, but because the signal is generically dominated by the small halos and subhalos, even a small uncertainty in these parameters can have a substantial impact on the boost. Changing the power law index $\alpha_h$ from $-1.05$ to $-1.0$, in the previous example, leaves the MW boost factor unaffected, but reduces the integrated boost factor for the diffuse gamma-rays to $\sim 30$, in which case $\Delta(8.5\kpc)/\mathrm{BF}_\mathrm{isotropic} \sim 1/5$ and $1/15$.

Raising the cutoff mass above 10$^{-6} M_\odot$ also improves the consistency with the constraints (at least for the default value of $\alpha_h = -1.05$), simply because most of the extragalactic signal comes from small, dense subhalos which are almost entirely destroyed if the cutoff is sufficiently raised. As an example, again taking $\alpha_\mathrm{sub} = -1.05$, $A_\mathrm{sub} = 10^{-0.2}$ but  
raising the cutoff mass to $1 M_\odot$ 
reduces the boosts for the MW and the diffuse emission to $\sim 15$ and $\sim 25$ respectively, yielding $\Delta(8.5\kpc)/\mathrm{BF}_\mathrm{isotropic} \sim 1/10$ and $1/30$. Raising the cutoff mass thus opens additional allowed parameter space at larger values of $\alpha_\mathrm{sub}$.

In summary, the diffuse extragalactic gamma-ray background is probably (together with the CMB) one of the most sensitive probes of the substructure-dominated scenario for the PAMELA excess.  The expected isotropic signals in such a case are typically of the same order as the current limits, but they are extremely sensitive to even small changes in the parametrization of substructure.  As such, the substructure-dominated scenario for the PAMELA excess does not appear to be in clear conflict with  \cite{Zavala:2011tt}, although it may still require a relatively small contribution from star forming galaxies and blazars to the gamma ray background. It is interesting to note that the naively least-constrained scenario, where the substructure is minimized, may not actually be least constrained as an explanation for PAMELA when local substructure is self-consistently taken into account.

\subsection{Depletion of substructure in light mediator models}

In the discussion above, we have assumed both that the DM can be adequately modeled as collisionless, and that the kinetic decoupling temperature (which sets the low-mass cutoff scale for substructure) is the same as in the standard WIMP scenario. However, in the presence of a light mediator, the DM \emph{does} possess a potentially non-negligible self-interaction, and the natural kinetic decoupling temperature can be substantially smaller. Here we briefly explore the possible effects of a light mediator on small-scale substructure.

The usually assumed low-mass cutoff of $\sim 10^{-6} M_\odot$ presumes a kinetic decoupling temperature of $\mathcal{O}(100)$ MeV, which is appropriate for a standard WIMP. For models with a light mediator kinetically mixed with the photon, however, the cross section for scattering of DM on charged SM particles can be much larger. Direct detection experiments constrain the mixing if the scattering is elastic, but small ($\mathcal{O}(100)$ keV) mass splittings $\delta$ between the states in the DM multiplet can remove these limits.

For plausible parameters, the DM may remain coupled to the SM via DM-electron scattering at temperatures down to $m_e$: specifically, the kinetic decoupling temperature is \cite{Feng:2010zp}, 
\begin{equation} T_\mathrm{kd}^e \sim \mathrm{max} \left\{m_e, \delta, 0.82 \mathrm{MeV} \left[\frac{10^{-3}}{\epsilon} \right]^{1/2} \times 
\left[\frac{m_\phi}{30 \mathrm{MeV}} \right]  \left[\frac{0.021}{\alpha_D} \right]^{1/4} \left[\frac{m_\chi}{\mathrm{TeV}} \right]^{1/4} \right\}. \end{equation}
For these relatively low kinetic decoupling temperatures, the mass cutoff scale is given by \cite{Bringmann:2009vf},
 \begin{equation} M_\mathrm{cutoff} = 3.4 \times 10^{-6} \left(\frac{T_\mathrm{kd} g_\mathrm{eff}^{1/4}}{50 \mathrm{MeV}} \right)^{-3}. \end{equation}
For $T_\mathrm{kd} \sim m_e$, we find $M_\mathrm{cutoff} \sim 1 M_\odot$. It is certainly not \emph{necessary} that the cutoff mass be this high, since small $\epsilon$ and $\alpha_D$ would lower the cutoff scale, but it is plausible for scenarios with very small mediator mass.

After structure formation, subhalos may also be evaporated in the presence of self-interaction, by collisions with more energetic particles in the host halos. This mechanism has already been invoked to set constraints on the self-interaction by demanding that dwarf galaxies within the MW and galaxies within clusters have not yet evaporated, following \cite{Buckley:2009in}, but what of subhalos in smaller, denser hosts? Would they evaporate early, for models saturating the self-interaction limits we have imposed?

Note that the properties of the \emph{subhalo} do not matter for this question, since the particles of the host halo are by definition not bound to substructure and have enough energy to remove particles from any subhalo. Only the characteristic velocity and density of the host halo are relevant. Consequently, this effect cannot strongly affect local $\Delta$ (since the MW has dwarf galaxy subhalos), and we need only ask if it can be relevant in smaller host halos, thus affecting the diffuse gamma-ray limits.

The timescale for evaporation scales as $(n \sigma(v) v)^{-1}$, where $v$ and $n$ are the characteristic velocity and number density of the host halo. If we use the parameterization of \cite{Kamionkowski:2010mi} and take $v \propto n^{-1.75}$, the evaporation timescale varies as $(v^{0.43} \sigma(v))^{-1}$. For small $v$, $\sigma$ scales as $v^{-0.7}$ before leveling off to log dependence on $v$ (Equation \ref{eq:sigmat}). Thus, while the dependence on $v$ is always quite weak, it seems possible in principle for the saturation of the self-interaction to pick out a particular range of halo masses in which evaporation is faster than the age of the universe, with halos above or below this characteristic mass range not experiencing subhalo evaporation. 

In general, for couplings smaller than the self-interaction threshold (where self-scattering is too weak to change halo properties), evaporation is never fast enough to efficiently destroy subhalos, at least using this simple estimate. For lighter force carriers, for which the DM departs from the collisionless limit, evaporation can naturally occur over some range of host halo masses. This evaporation could further weaken diffuse isotropic gamma-ray limits, which under the usual assumptions receive large contributions from the substructure of low-mass halos. Moreover, even in the nominally ``collisionless'' region of parameter space, the gap between the timescale for evaporation and the age of the halo can be less than an order of magnitude, and a more careful analysis is justified.

\section{Conclusions}
The $e^+$ excess observed by PAMELA and now \emph{Fermi} points to a new primary source of cosmic ray electrons and positrons. One of the most exciting, if speculative, explanations of the excess is that it arises from Sommerfeld-enhanced DM annihilation. Such models naturally provide annihilation rates much larger than would be expected from a thermal WIMP with substructure enhancement alone.

Nonetheless, in models with Sommerfeld enhancement, in the presence of $\mathcal{O}(1)$ substructure, the substructure is often the {\em dominant} source of the signal, because of the low velocity dispersion of the bound subhalos. 
Most constraints on these models have been calculated assuming the signal arises from the smooth halo, and the limits become dramatically weaker if substructure dominates the CR signal. 

In particular, the constraints on parameter space from the CMB are removed for $\Delta \gtrsim 0.4$, since the local signal as well as the early universe signal are both determined by the saturated cross section, in contrast to the smooth halo piece that is generally unsaturated. Because substructure is depleted in the inner regions of galaxies, constraints from FSR and ICS signals in the inner MW are strongly suppressed. As $\Delta$ increases, lower couplings between the DM and the force carrier are required to fit the CR excesses, and this in turn relaxes limits on the force carrier mass from bounds on DM self-interaction; such constraints are subsumed by the CMB bounds, but depending on the model, the more stringent requirement that self-interaction have negligible effect on dwarf galaxy structure may imply $m_\phi \gtrsim 10$ MeV. This has profound implications for terrestrial searches. Including the effects of substructure not only opens the low-mass ($m_\phi \sim 1-200 \mev$) region of parameter space, but possibly makes it the preferred range, once additional constraints are considered. Given the sensitivity that many experiments have in this region, these searches become even more motivated.

Limits on the DM annihilation rate from measurements of the diffuse extragalactic gamma-ray background, or from gamma-rays from the outer halo of the MW, generally become stronger as the amount of substructure is increased. However, the key quantity is the ratio of the  signal in these constraining channels to the local substructure-enhanced annihilation rate, and this can easily be substantially \emph{smaller} at higher $\Delta$ than for $\Delta=0$, relaxing the limits on DM explanations for the CR excesses.

There is no established consensus on what the local boost from substructure should be, due to the large uncertainties in extrapolating the results from $N$-body simulations below their mass resolutions, but it is not thought to be extremely large. Nonetheless, in the presence of Sommerfeld enhancement the substructure contribution can still easily dominate and open up new regions of parameter space, especially for sub-200 MeV force carriers. These regions, too, will be tested by further observations, with the exciting possibility that we might already be detecting the cosmic ray signals of DM substructure.

\acknowledgments

We thank Nima Arkani-Hamed, Rouven Essig, Mike Kuhlen, Julien Lavalle, Annika Peter, Jennifer Siegal-Gaskins, and 
Jes\'us Zavala for comments and conversations. NW is supported by NSF grant \#0947827, as well as support from the Ambrose Monell Foundation. TS is supported by NSF grant AST-0807444 and DOE grant DE-FG02-90ER40542. Research at the Perimeter Institute is supported in part by the Government of Canada through NSERC and by the Province of Ontario through MEDT.  NW and TS acknowledge the hospitality of the Aspen Center for Theoretical Physics during the early stages of this work. This research was supported in part by the National Science Foundation under Grant No. NSF PHY05-51164.

\begin{appendix}
\section{An inelastic example}
\label{app:inelastic}

\begin{figure*}
\includegraphics[width=0.49\textwidth]{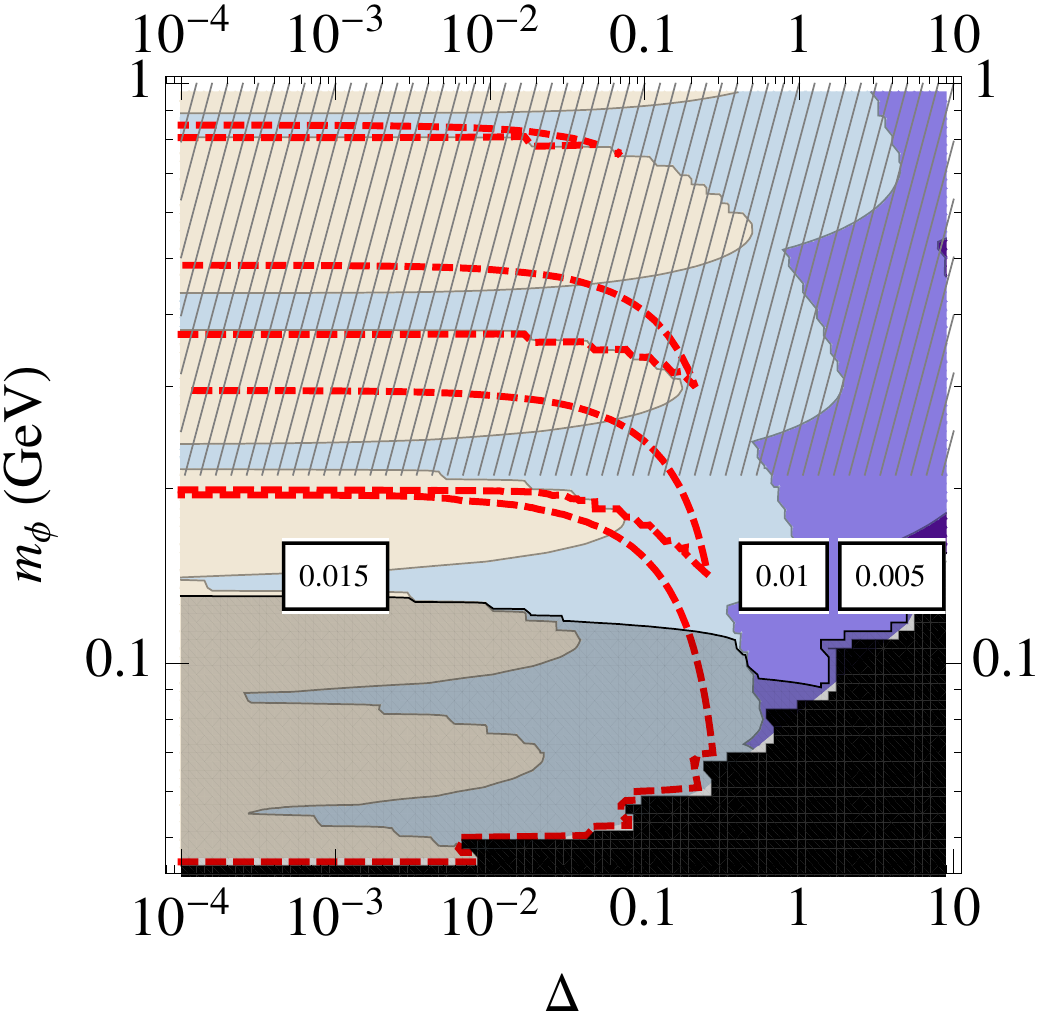}
\includegraphics[width=0.45\textwidth]{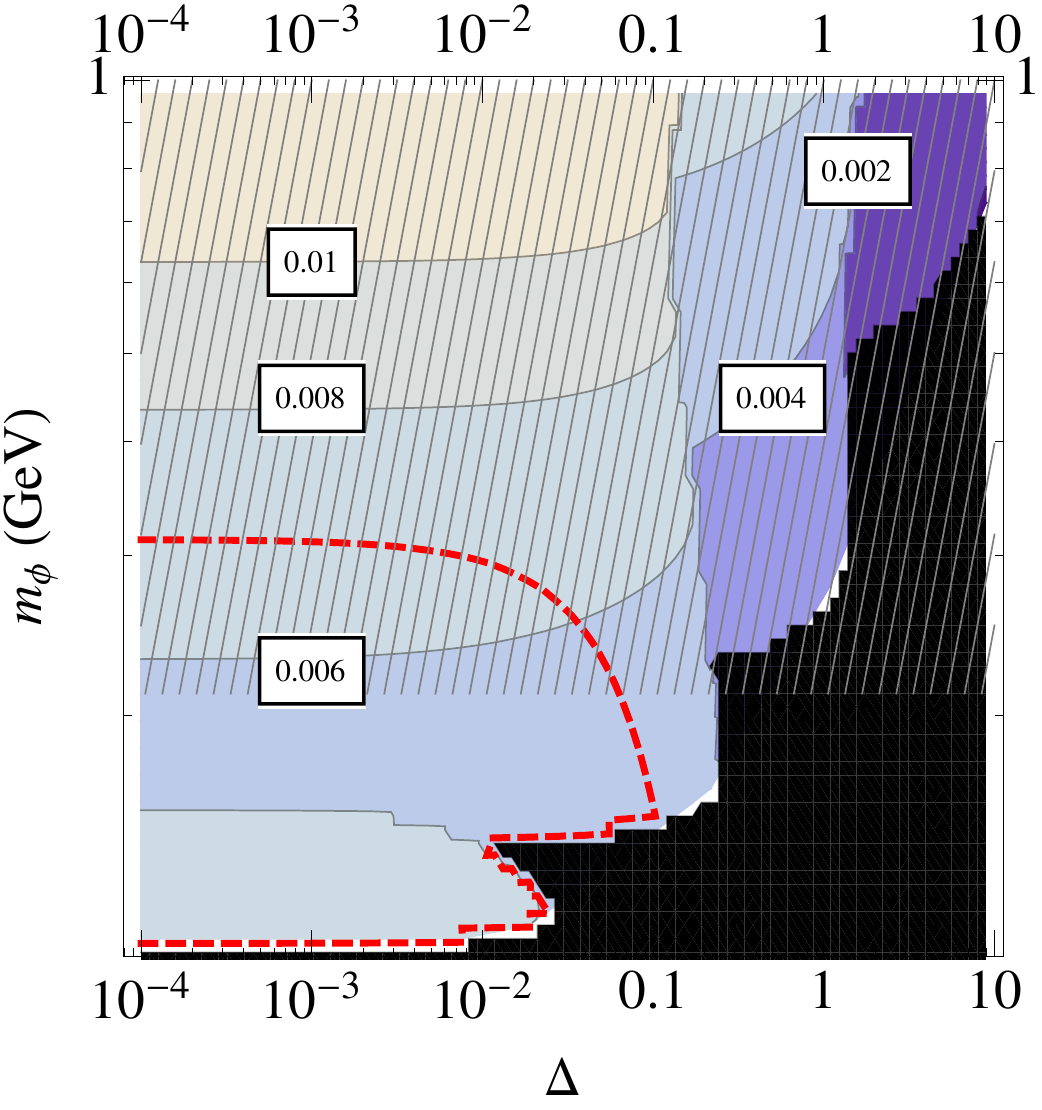}
\caption{The dark sector coupling $\alpha_D$ as a function of mediator mass $m_\phi$ and substructure contribution $\Delta$, for a fixed DM mass of 1 TeV, mass splitting of 700 keV, and local aggregate boost factor of 65, in scenarios 1 (\emph{left panel}) and 2 (\emph{right panel}); see text for descriptions of the two scenarios. Regions to the left of and/or below the red dashed line are ruled out by constraints from the CMB; the self-interaction bounds lie at mediator masses below the range of this plot. In the regions overlaid in solid black, the approximation we have used for the multi-state Sommerfeld enhancement is expected to break down; in the grayed-out regions, the model is unphysical as the required cross section to obtain the correct relic density is smaller than the minimal contribution from $t$-channel annihilation into dark gauge bosons. The dark gauge boson is assumed to decay into electrons only, in which case this boost factor and DM mass provide a good fit to the PAMELA and \emph{Fermi} data. When the gauge boson mass exceeds twice the muon mass, the true final state may become more complicated, so this region is indicated by cross-hatching.}
\label{fig:inelastic_alpha}
\end{figure*}

\begin{figure*}
\includegraphics[width=0.49\textwidth]{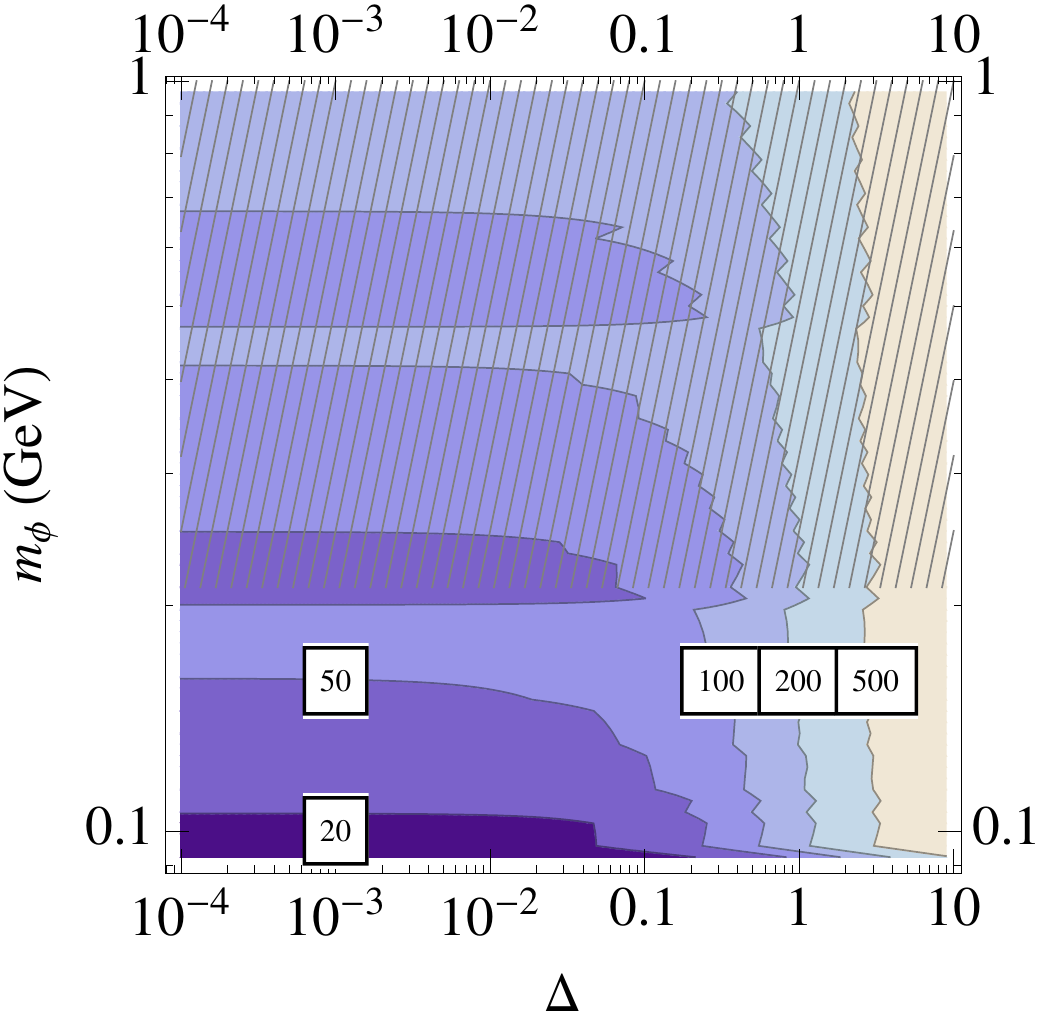}
\includegraphics[width=0.45\textwidth]{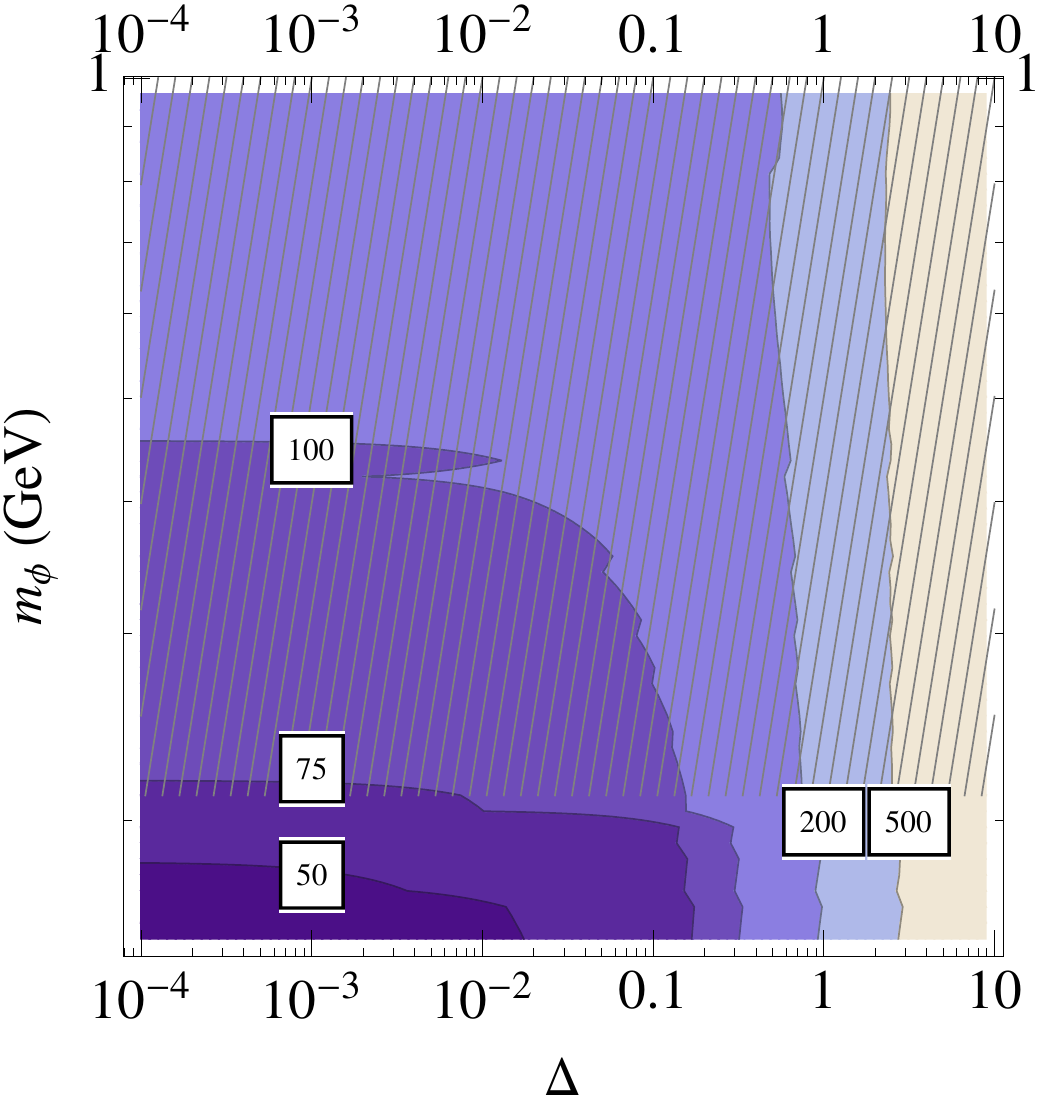}
\caption{The maximum local boost factor for 1 TeV DM with a 700 keV mass splitting, consistent with constraints from the thermal relic density, the CMB, self-interaction bounds, and naturalness (in the sense of not relying on the resonance peaks), in scenarios 1 (\emph{left panel}) and 2 (\emph{right panel}). The dark gauge boson is assumed to decay into electrons only; when the gauge boson mass exceeds twice the muon mass, the true final state may become more complicated, so this region is indicated by cross-hatching. The regions overlaid in solid black indicate where our approximation for the multi-state Sommerfeld enhancement is expected to break down.}
\label{fig:inelastic_maxlocalboost}
\end{figure*}

The observant reader may have noticed that in Fig. \ref{fig:compareconstraints}, the zero-$\Delta$ case appears to be ruled out for masses up to 1 GeV by the CMB bounds, and in Fig. \ref{fig:maxlocalboost}, the maximum local boost for $\Delta=0$ is only $\sim 30$ at $m_\phi = 200$ MeV. At first glance, this seems to contradict the claim that the PAMELA and \emph{Fermi} results can be explained by Sommerfeld-enhanced models in the absence of substructure, for sub-GeV mediators.

However, this is just the factor of $\sim 2-3$ discrepancy noted in \cite{Finkbeiner:2010sm}, for the case where the states in the dark-charged DM multiplet were taken to be degenerate. As briefly mentioned in Sec. \ref{subsec:relicdensity}, the benchmark models presented in \cite{Finkbeiner:2010sm} possessed two non-degenerate DM states, and a larger smooth-halo Sommerfeld enhancement as a result. The same analysis described above can be applied to a model of this type with a fixed mass splitting, with the modification that now we also need to take into account annihilation channels involving the excited states, at least for the freezeout calculation. 

One example of a Lagrangian with the desired features takes the form:
\begin{align} \mathcal{L} & =  i \bar{\Psi} \gamma^\mu \left( \partial_\mu + i g_D \phi_\mu \right) \Psi + \left(\partial^\mu + i g_D \phi^\mu \right) h_D \left(\partial_\mu - i g_D \phi_\mu \right) h_D^* - m_\chi \bar{\Psi} \Psi - \frac{y}{2 \Lambda} \left( \bar{\Psi^C} \Psi h_D^* h_D^* + \bar{\Psi} \Psi^C h_D h_D \right) \nonumber \\ &  -\frac{1}{4} F^\mathrm{D}_{\mu \nu} F_D^{\mu \nu} - \frac{\epsilon}{2} F^\mathrm{EM}_{\mu \nu} F_D^{\mu\nu} + V(|h_D|^2) + \mathcal{L}_\mathrm{SM}, \end{align}
above the symmetry breaking scale, where the DM is the Dirac fermion $\Psi$, $h_D$ is a dark Abelian Higgs, and $\Lambda$ is a high-energy scale associated with the dimension-5 operator. Below the symmetry breaking scale, this operator confers a Majorana mass on the DM: going to unitary gauge and writing $h_D \rightarrow (v_D + \rho)/\sqrt{2}$, we can rewrite the Lagrangian in terms of the mass-eigenstate Majorana fermions $\chi_1, \chi_2$ (see Sec. \ref{sec:lagrangian}),
\begin{align} \mathcal{L} & =  i  \left(\chi_1^\dagger \quad \chi_2^\dagger \right)  \left(\begin{array}{cc} \bar{\sigma}^\mu \partial_\mu & - g_D  \bar{\sigma}^\mu \phi_\mu \\ g_D  \bar{\sigma}^\mu \phi_\mu &  \bar{\sigma}^\mu \partial_\mu \end{array} \right) \left(\begin{array}{c} \chi_1 \\ \chi_2 \end{array} \right) + \frac{m_\phi^2}{2} \phi^\mu \phi_\mu + \frac{g_D^2}{2} \rho \rho \phi^\mu \phi_\mu + v_D g_D^2 \rho \phi^\mu \phi_\mu + \frac{1}{2} \partial^\mu \rho \partial_\mu \rho \nonumber \\ &  - \frac{y}{4 \Lambda} (\chi_1 \chi_1 - \chi_2 \chi_2) (\rho^2 + 2 v_D \rho) - \frac{1}{2} \left(\chi_1 \quad \chi_2 \right)  \left(\begin{array}{cc} m_\chi + m_M & 0 \\ 0 & m_\chi - m_M \end{array} \right) \left(\begin{array}{c} \chi_1 \\ \chi_2 \end{array} \right) + h.c. \nonumber \\ &  -\frac{1}{4} F^\mathrm{D}_{\mu \nu} F_D^{\mu \nu} - \frac{\epsilon}{2} F^\mathrm{EM}_{\mu \nu} F_D^{\mu\nu} + V(v_D, \rho) + \mathcal{L}_\mathrm{SM}, \label{eq:inelasticlagrangian} \end{align}
where $m_M = y v_D^2/ 2 \Lambda$ and $m_\phi = g_D v_D$.

The DM can now annihilate into both the dark gauge bosons and the dark Higgs; however, as explained in detail in  \cite{Finkbeiner:2010sm}, the annihilation into final states involving the dark Higgs $\rho$ is suppressed at late times (the $s$-wave annihilation through an intermediate $\phi$ boson requires one of the initial DM particles to be in the excited state $\chi_2$, which generally has a lifetime shorter than the age of the universe; the direct annihilation through the $\chi_1 \chi_1 \rho^2$ operator is strongly suppressed at low velocities). Thus, these annihilation channels are ``irrelevant'' by the definition given in \S \ref{subsec:relicdens}.

We again consider two scenarios: in our new scenario (1), for our ``baseline'' annihilation cross section we include all the annihilation channels that follow from the Lagrangian of Equation \ref{eq:inelasticlagrangian}; a point in parameter space is ruled out by the relic density constraint if this baseline cross section over-depletes the relic density. We assume that additional ``irrelevant'' (to indirect detection) annihilation channels can exist to deplete the relic density: this is easily achieved by e.g. introducing new states charged under the dark-sector gauge interaction, since $s$-channel annihilation through an off-shell dark gauge boson provides a coannihilation channel which is large at freezeout but suppressed by the abundance of the excited state in the present day.

In the inelastic scenario (2), rather than use the specific model described above (and taken from \cite{Finkbeiner:2010sm}), we assume a single universal $s$-wave annihilation cross section for co-annihilation and self-annihilation of DM particles in the ground or excited states, choose this cross section to obtain the correct relic density, and assume this annihilation rate is multiplied by the Sommerfeld enhancement. A point in parameter space is ``ruled out by the relic density constraint'' if the required cross section is smaller than $\sigma v = \pi \alpha_D^2 / m_\chi^2$. This approach removes differences between the inelastic and elastic cases due to differences in the self-annihilation vs co-annihilation rates and model-dependent extra annihilation channels, allowing us to clearly see the effect of the modified Sommerfeld enhancement in inelastic models.

We employ the approximation for the inelastic Sommerfeld enhancement derived by \cite{Slatyer:2009vg}. This approximation is expected to break down for mass splittings $\delta \gtrsim \alpha_D m_\phi$: consequently, we cannot study the small-mediator-mass region in detail in the inelastic case, especially in scenario (2) where at small mediator masses and/or large $\Delta$ very small values of $\alpha_D$ are favored.

Our results are shown in Figs. \ref{fig:inelastic_alpha}-\ref{fig:inelastic_maxlocalboost}: in Fig. \ref{fig:inelastic_alpha}, we take a target boost factor of 65, a mass splitting of 700 keV, and a DM mass of 1 TeV, motivated by the lowest-mediator-mass benchmark model in \cite{Finkbeiner:2010sm}. The qualitative features are similar to the elastic case, although we cannot study the self-interaction bounds since they only apply to small mediator masses where our approximations are expected to break down. We note that as expected, a local boost of 65 is permitted for a 200 MeV mediator and 1 TeV DM in the zero-$\Delta$ limit.

\end{appendix}
\bibliography{sommerfeldsubstructure}

\begin{thebibliography}{91}
\expandafter\ifx\csname natexlab\endcsname\relax\def\natexlab#1{#1}\fi
\expandafter\ifx\csname bibnamefont\endcsname\relax
  \def\bibnamefont#1{#1}\fi
\expandafter\ifx\csname bibfnamefont\endcsname\relax
  \def\bibfnamefont#1{#1}\fi
\expandafter\ifx\csname citenamefont\endcsname\relax
  \def\citenamefont#1{#1}\fi
\expandafter\ifx\csname url\endcsname\relax
  \def\url#1{\texttt{#1}}\fi
\expandafter\ifx\csname urlprefix\endcsname\relax\def\urlprefix{URL }\fi
\providecommand{\bibinfo}[2]{#2}
\providecommand{\eprint}[2][]{\url{#2}}

\bibitem[{\citenamefont{Adriani et~al.}(2009)}]{Adriani:2008zr}
\bibinfo{author}{\bibfnamefont{O.}~\bibnamefont{Adriani}} \bibnamefont{et~al.}
  (\bibinfo{collaboration}{PAMELA}), \bibinfo{journal}{Nature}
  \textbf{\bibinfo{volume}{458}}, \bibinfo{pages}{607} (\bibinfo{year}{2009}),
  \eprint{0810.4995}.

\bibitem[{\citenamefont{Abdo et~al.}(2009)}]{Abdo:2009zk}
\bibinfo{author}{\bibfnamefont{A.~A.} \bibnamefont{Abdo}} \bibnamefont{et~al.}
  (\bibinfo{collaboration}{The Fermi LAT}), \bibinfo{journal}{Phys. Rev. Lett.}
  \textbf{\bibinfo{volume}{102}}, \bibinfo{pages}{181101}
  (\bibinfo{year}{2009}), \eprint{0905.0025}.

\bibitem[{\citenamefont{Ackermann et~al.}(2010)}]{Ackermann:2010ij}
\bibinfo{author}{\bibfnamefont{M.}~\bibnamefont{Ackermann}}
  \bibnamefont{et~al.} (\bibinfo{collaboration}{Fermi LAT}),
  \bibinfo{journal}{Phys. Rev.} \textbf{\bibinfo{volume}{D82}},
  \bibinfo{pages}{092004} (\bibinfo{year}{2010}), \eprint{1008.3999}.

\bibitem[{\citenamefont{Chang et~al.}(2008)}]{aticlatest}
\bibinfo{author}{\bibfnamefont{J.}~\bibnamefont{Chang}} \bibnamefont{et~al.},
  \bibinfo{journal}{Nature} \textbf{\bibinfo{volume}{456}},
  \bibinfo{pages}{362} (\bibinfo{year}{2008}).

\bibitem[{\citenamefont{Panov et~al.}(2011)}]{Panov:2011zw}
\bibinfo{author}{\bibfnamefont{A.~D.} \bibnamefont{Panov}}
  \bibnamefont{et~al.}, \bibinfo{journal}{Astrophys. Space Sci. Trans.}
  \textbf{\bibinfo{volume}{7}}, \bibinfo{pages}{119} (\bibinfo{year}{2011}),
  \eprint{1104.3452}.

\bibitem[{\citenamefont{Ackermann et~al.}(2012)}]{FermiLAT:2011ab}
\bibinfo{author}{\bibfnamefont{M.}~\bibnamefont{Ackermann}}
  \bibnamefont{et~al.} (\bibinfo{collaboration}{The Fermi LAT Collaboration}),
  \bibinfo{journal}{Phys.Rev.Lett.} \textbf{\bibinfo{volume}{108}},
  \bibinfo{pages}{011103} (\bibinfo{year}{2012}), \bibinfo{note}{5 figures, 1
  table, revtex 4.1}, \eprint{1109.0521}.

\bibitem[{\citenamefont{Finkbeiner and Weiner}(2007)}]{Finkbeiner:2007kk}
\bibinfo{author}{\bibfnamefont{D.~P.} \bibnamefont{Finkbeiner}}
  \bibnamefont{and} \bibinfo{author}{\bibfnamefont{N.}~\bibnamefont{Weiner}},
  \bibinfo{journal}{Phys. Rev.} \textbf{\bibinfo{volume}{D76}},
  \bibinfo{pages}{083519} (\bibinfo{year}{2007}), \eprint{astro-ph/0702587}.

\bibitem[{\citenamefont{Arkani-Hamed et~al.}(2009)\citenamefont{Arkani-Hamed,
  Finkbeiner, Slatyer, and Weiner}}]{ArkaniHamed:2008qn}
\bibinfo{author}{\bibfnamefont{N.}~\bibnamefont{Arkani-Hamed}},
  \bibinfo{author}{\bibfnamefont{D.~P.} \bibnamefont{Finkbeiner}},
  \bibinfo{author}{\bibfnamefont{T.~R.} \bibnamefont{Slatyer}},
  \bibnamefont{and} \bibinfo{author}{\bibfnamefont{N.}~\bibnamefont{Weiner}},
  \bibinfo{journal}{Phys. Rev.} \textbf{\bibinfo{volume}{D79}},
  \bibinfo{pages}{015014} (\bibinfo{year}{2009}), \eprint{0810.0713}.

\bibitem[{\citenamefont{Pospelov and Ritz}(2009)}]{Pospelov:2008jd}
\bibinfo{author}{\bibfnamefont{M.}~\bibnamefont{Pospelov}} \bibnamefont{and}
  \bibinfo{author}{\bibfnamefont{A.}~\bibnamefont{Ritz}},
  \bibinfo{journal}{Phys. Lett.} \textbf{\bibinfo{volume}{B671}},
  \bibinfo{pages}{391} (\bibinfo{year}{2009}), \eprint{0810.1502}.

\bibitem[{\citenamefont{Finkbeiner et~al.}(2010)\citenamefont{Finkbeiner,
  Goodenough, Slatyer, Vogelsberger, and Weiner}}]{Finkbeiner:2010sm}
\bibinfo{author}{\bibfnamefont{D.~P.} \bibnamefont{Finkbeiner}},
  \bibinfo{author}{\bibfnamefont{L.}~\bibnamefont{Goodenough}},
  \bibinfo{author}{\bibfnamefont{T.~R.} \bibnamefont{Slatyer}},
  \bibinfo{author}{\bibfnamefont{M.}~\bibnamefont{Vogelsberger}},
  \bibnamefont{and} \bibinfo{author}{\bibfnamefont{N.}~\bibnamefont{Weiner}}
  (\bibinfo{year}{2010}), \eprint{1011.3082}.

\bibitem[{\citenamefont{Lattanzi and Silk}(2009)}]{Lattanzi:2008qa}
\bibinfo{author}{\bibfnamefont{M.}~\bibnamefont{Lattanzi}} \bibnamefont{and}
  \bibinfo{author}{\bibfnamefont{J.~I.} \bibnamefont{Silk}},
  \bibinfo{journal}{Phys. Rev.} \textbf{\bibinfo{volume}{D79}},
  \bibinfo{pages}{083523} (\bibinfo{year}{2009}), \eprint{0812.0360}.

\bibitem[{\citenamefont{Kuhlen and Malyshev}(2009)}]{Kuhlen:2009is}
\bibinfo{author}{\bibfnamefont{M.}~\bibnamefont{Kuhlen}} \bibnamefont{and}
  \bibinfo{author}{\bibfnamefont{D.}~\bibnamefont{Malyshev}},
  \bibinfo{journal}{Phys. Rev.} \textbf{\bibinfo{volume}{D79}},
  \bibinfo{pages}{123517} (\bibinfo{year}{2009}), \eprint{0904.3378}.

\bibitem[{\citenamefont{Kuhlen et~al.}(2009)\citenamefont{Kuhlen, Madau, and
  Silk}}]{Kuhlen:2009kx}
\bibinfo{author}{\bibfnamefont{M.}~\bibnamefont{Kuhlen}},
  \bibinfo{author}{\bibfnamefont{P.}~\bibnamefont{Madau}}, \bibnamefont{and}
  \bibinfo{author}{\bibfnamefont{J.}~\bibnamefont{Silk}},
  \bibinfo{journal}{Science} \textbf{\bibinfo{volume}{325}},
  \bibinfo{pages}{970} (\bibinfo{year}{2009}), \eprint{0907.0005}.

\bibitem[{\citenamefont{Bovy}(2009)}]{Bovy:2009zs}
\bibinfo{author}{\bibfnamefont{J.}~\bibnamefont{Bovy}}, \bibinfo{journal}{Phys.
  Rev.} \textbf{\bibinfo{volume}{D79}}, \bibinfo{pages}{083539}
  (\bibinfo{year}{2009}), \eprint{0903.0413}.

\bibitem[{\citenamefont{Yuan et~al.}(2009)}]{Yuan:2009bb}
\bibinfo{author}{\bibfnamefont{Q.}~\bibnamefont{Yuan}} \bibnamefont{et~al.},
  \bibinfo{journal}{JCAP} \textbf{\bibinfo{volume}{0912}}, \bibinfo{pages}{011}
  (\bibinfo{year}{2009}), \eprint{0905.2736}.

\bibitem[{\citenamefont{Vincent et~al.}(2010)\citenamefont{Vincent, Xue, and
  Cline}}]{Vincent:2010kv}
\bibinfo{author}{\bibfnamefont{A.~C.} \bibnamefont{Vincent}},
  \bibinfo{author}{\bibfnamefont{W.}~\bibnamefont{Xue}}, \bibnamefont{and}
  \bibinfo{author}{\bibfnamefont{J.~M.} \bibnamefont{Cline}}
  (\bibinfo{year}{2010}), \eprint{1009.5383}.

\bibitem[{\citenamefont{Kistler and Siegal-Gaskins}(2010)}]{Kistler:2009xf}
\bibinfo{author}{\bibfnamefont{M.~D.} \bibnamefont{Kistler}} \bibnamefont{and}
  \bibinfo{author}{\bibfnamefont{J.~M.} \bibnamefont{Siegal-Gaskins}},
  \bibinfo{journal}{Phys. Rev.} \textbf{\bibinfo{volume}{D81}},
  \bibinfo{pages}{103521} (\bibinfo{year}{2010}), \eprint{0909.0519}.

\bibitem[{\citenamefont{Kamionkowski et~al.}(2010)\citenamefont{Kamionkowski,
  Koushiappas, and Kuhlen}}]{Kamionkowski:2010mi}
\bibinfo{author}{\bibfnamefont{M.}~\bibnamefont{Kamionkowski}},
  \bibinfo{author}{\bibfnamefont{S.~M.} \bibnamefont{Koushiappas}},
  \bibnamefont{and} \bibinfo{author}{\bibfnamefont{M.}~\bibnamefont{Kuhlen}},
  \bibinfo{journal}{Phys. Rev.} \textbf{\bibinfo{volume}{D81}},
  \bibinfo{pages}{043532} (\bibinfo{year}{2010}), \eprint{1001.3144}.

\bibitem[{\citenamefont{Pospelov}(2009)}]{Pospelov:2008zw}
\bibinfo{author}{\bibfnamefont{M.}~\bibnamefont{Pospelov}},
  \bibinfo{journal}{Phys.Rev.} \textbf{\bibinfo{volume}{D80}},
  \bibinfo{pages}{095002} (\bibinfo{year}{2009}), \eprint{0811.1030}.

\bibitem[{\citenamefont{Batell et~al.}(2009{\natexlab{a}})\citenamefont{Batell,
  Pospelov, and Ritz}}]{Batell:2009yf}
\bibinfo{author}{\bibfnamefont{B.}~\bibnamefont{Batell}},
  \bibinfo{author}{\bibfnamefont{M.}~\bibnamefont{Pospelov}}, \bibnamefont{and}
  \bibinfo{author}{\bibfnamefont{A.}~\bibnamefont{Ritz}},
  \bibinfo{journal}{Phys. Rev.} \textbf{\bibinfo{volume}{D79}},
  \bibinfo{pages}{115008} (\bibinfo{year}{2009}{\natexlab{a}}),
  \eprint{0903.0363}.

\bibitem[{\citenamefont{Essig et~al.}(2009{\natexlab{a}})\citenamefont{Essig,
  Schuster, and Toro}}]{Essig:2009nc}
\bibinfo{author}{\bibfnamefont{R.}~\bibnamefont{Essig}},
  \bibinfo{author}{\bibfnamefont{P.}~\bibnamefont{Schuster}}, \bibnamefont{and}
  \bibinfo{author}{\bibfnamefont{N.}~\bibnamefont{Toro}},
  \bibinfo{journal}{Phys. Rev.} \textbf{\bibinfo{volume}{D80}},
  \bibinfo{pages}{015003} (\bibinfo{year}{2009}{\natexlab{a}}),
  \eprint{0903.3941}.

\bibitem[{\citenamefont{Reece and Wang}(2009)}]{Reece:2009un}
\bibinfo{author}{\bibfnamefont{M.}~\bibnamefont{Reece}} \bibnamefont{and}
  \bibinfo{author}{\bibfnamefont{L.-T.} \bibnamefont{Wang}},
  \bibinfo{journal}{JHEP} \textbf{\bibinfo{volume}{07}}, \bibinfo{pages}{051}
  (\bibinfo{year}{2009}).

\bibitem[{\citenamefont{Aubert et~al.}(2009)}]{:2009pw}
\bibinfo{author}{\bibfnamefont{B.}~\bibnamefont{Aubert}} \bibnamefont{et~al.}
  (\bibinfo{collaboration}{BABAR Collaboration}) (\bibinfo{year}{2009}),
  \eprint{0908.2821}.

\bibitem[{\citenamefont{Bjorken et~al.}(2009)\citenamefont{Bjorken, Essig,
  Schuster, and Toro}}]{Bjorken:2009mm}
\bibinfo{author}{\bibfnamefont{J.~D.} \bibnamefont{Bjorken}},
  \bibinfo{author}{\bibfnamefont{R.}~\bibnamefont{Essig}},
  \bibinfo{author}{\bibfnamefont{P.}~\bibnamefont{Schuster}}, \bibnamefont{and}
  \bibinfo{author}{\bibfnamefont{N.}~\bibnamefont{Toro}},
  \bibinfo{journal}{Phys. Rev.} \textbf{\bibinfo{volume}{D80}},
  \bibinfo{pages}{075018} (\bibinfo{year}{2009}), \eprint{0906.0580}.

\bibitem[{\citenamefont{Batell et~al.}(2009{\natexlab{b}})\citenamefont{Batell,
  Pospelov, and Ritz}}]{Batell:2009di}
\bibinfo{author}{\bibfnamefont{B.}~\bibnamefont{Batell}},
  \bibinfo{author}{\bibfnamefont{M.}~\bibnamefont{Pospelov}}, \bibnamefont{and}
  \bibinfo{author}{\bibfnamefont{A.}~\bibnamefont{Ritz}},
  \bibinfo{journal}{Phys. Rev.} \textbf{\bibinfo{volume}{D80}},
  \bibinfo{pages}{095024} (\bibinfo{year}{2009}{\natexlab{b}}),
  \eprint{0906.5614}.

\bibitem[{\citenamefont{Bergsma et~al.}(1985)}]{Bergsma:1985qz}
\bibinfo{author}{\bibfnamefont{F.}~\bibnamefont{Bergsma}} \bibnamefont{et~al.}
  (\bibinfo{collaboration}{CHARM}), \bibinfo{journal}{Phys. Lett.}
  \textbf{\bibinfo{volume}{B157}}, \bibinfo{pages}{458} (\bibinfo{year}{1985}).

\bibitem[{\citenamefont{Riordan et~al.}(1987)}]{Riordan:1987aw}
\bibinfo{author}{\bibfnamefont{E.~M.} \bibnamefont{Riordan}}
  \bibnamefont{et~al.}, \bibinfo{journal}{Phys. Rev. Lett.}
  \textbf{\bibinfo{volume}{59}}, \bibinfo{pages}{755} (\bibinfo{year}{1987}).

\bibitem[{\citenamefont{Bjorken et~al.}(1988)}]{Bjorken:1988as}
\bibinfo{author}{\bibfnamefont{J.~D.} \bibnamefont{Bjorken}}
  \bibnamefont{et~al.}, \bibinfo{journal}{Phys. Rev.}
  \textbf{\bibinfo{volume}{D38}}, \bibinfo{pages}{3375} (\bibinfo{year}{1988}).

\bibitem[{\citenamefont{Bross et~al.}(1991)}]{Bross:1989mp}
\bibinfo{author}{\bibfnamefont{A.}~\bibnamefont{Bross}} \bibnamefont{et~al.},
  \bibinfo{journal}{Phys. Rev. Lett.} \textbf{\bibinfo{volume}{67}},
  \bibinfo{pages}{2942} (\bibinfo{year}{1991}).

\bibitem[{\citenamefont{Andreas and Ringwald}(2010)}]{Andreas:2010tp}
\bibinfo{author}{\bibfnamefont{S.}~\bibnamefont{Andreas}} \bibnamefont{and}
  \bibinfo{author}{\bibfnamefont{A.}~\bibnamefont{Ringwald}}
  (\bibinfo{year}{2010}), \eprint{arXiv:1008.4519}.

\bibitem[{\citenamefont{Archilli et~al.}(2011)}]{Archilli:2011nh}
\bibinfo{author}{\bibfnamefont{F.}~\bibnamefont{Archilli}} \bibnamefont{et~al.}
  (\bibinfo{year}{2011}), \eprint{arXiv:1107.2531}.

\bibitem[{\citenamefont{Merkel et~al.}(2011)}]{Merkel:2011ze}
\bibinfo{author}{\bibfnamefont{H.}~\bibnamefont{Merkel}} \bibnamefont{et~al.}
  (\bibinfo{collaboration}{A1}), \bibinfo{journal}{Phys. Rev. Lett.}
  \textbf{\bibinfo{volume}{106}}, \bibinfo{pages}{251802}
  (\bibinfo{year}{2011}).

\bibitem[{\citenamefont{Abrahamyan et~al.}(2011)}]{Abrahamyan:2011gv}
\bibinfo{author}{\bibfnamefont{S.}~\bibnamefont{Abrahamyan}}
  \bibnamefont{et~al.} (\bibinfo{year}{2011}), \eprint{1108.2750}.

\bibitem[{\citenamefont{Essig et~al.}(2009{\natexlab{b}})\citenamefont{Essig,
  Schuster, Toro, Wojtsekhowski et~al.}}]{Proposal}
\bibinfo{author}{\bibfnamefont{R.}~\bibnamefont{Essig}},
  \bibinfo{author}{\bibfnamefont{P.}~\bibnamefont{Schuster}},
  \bibinfo{author}{\bibfnamefont{N.}~\bibnamefont{Toro}},
  \bibinfo{author}{\bibfnamefont{B.}~\bibnamefont{Wojtsekhowski}},
  \bibnamefont{et~al.}, \bibinfo{journal}{JLab Experiment E12-10-009}
  (\bibinfo{year}{2009}{\natexlab{b}}).

\bibitem[{\citenamefont{Essig et~al.}(2011)\citenamefont{Essig, Schuster, Toro,
  and Wojtsekhowski}}]{Essig:2010xa}
\bibinfo{author}{\bibfnamefont{R.}~\bibnamefont{Essig}},
  \bibinfo{author}{\bibfnamefont{P.}~\bibnamefont{Schuster}},
  \bibinfo{author}{\bibfnamefont{N.}~\bibnamefont{Toro}}, \bibnamefont{and}
  \bibinfo{author}{\bibfnamefont{B.}~\bibnamefont{Wojtsekhowski}},
  \bibinfo{journal}{JHEP} \textbf{\bibinfo{volume}{02}}, \bibinfo{pages}{009}
  (\bibinfo{year}{2011}).

\bibitem[{\citenamefont{{\href{http://www.jlab.org/exp\_prog/PACpage/PAC37/pro%
posals/Proposals/New\%20Proposals/PR-11-006.pdf}{Jefferson Lab PAC37 Proposal
  PR-11-006.}}}()}]{HPS}
\bibinfo{author}{\bibnamefont{{\href{http://www.jlab.org/exp\_prog/PACpage/PAC%
37/proposals/Proposals/New\%20Proposals/PR-11-006.pdf}{Jefferson Lab PAC37
  Proposal PR-11-006.}}}}

\bibitem[{\citenamefont{Wojtsekhowski}(2009)}]{Wojtsekhowski:2009vz}
\bibinfo{author}{\bibfnamefont{B.}~\bibnamefont{Wojtsekhowski}},
  \bibinfo{journal}{AIP Conf. Proc.} \textbf{\bibinfo{volume}{1160}},
  \bibinfo{pages}{149} (\bibinfo{year}{2009}).

\bibitem[{\citenamefont{Freytsis et~al.}(2010)\citenamefont{Freytsis,
  Ovanesyan, and Thaler}}]{Freytsis:2009bh}
\bibinfo{author}{\bibfnamefont{M.}~\bibnamefont{Freytsis}},
  \bibinfo{author}{\bibfnamefont{G.}~\bibnamefont{Ovanesyan}},
  \bibnamefont{and} \bibinfo{author}{\bibfnamefont{J.}~\bibnamefont{Thaler}},
  \bibinfo{journal}{JHEP} \textbf{\bibinfo{volume}{01}}, \bibinfo{pages}{111}
  (\bibinfo{year}{2010}).

\bibitem[{\citenamefont{{e.g.
  \href{http://www.desy.de/\~ringwald/axions/talks/bim\_dark\_follow.pdf}{http%
://www.desy.de/\textasciitilde
  ringwald/axions/talks/bim\_dark\_follow.pdf.}}}()}]{HIPS}
\bibinfo{author}{\bibnamefont{{e.g.
  \href{http://www.desy.de/\~ringwald/axions/talks/bim\_dark\_follow.pdf}{http%
://www.desy.de/\textasciitilde ringwald/axions/talks/bim\_dark\_follow.pdf.}}}}

\bibitem[{\citenamefont{Dreiner et~al.}(2008)\citenamefont{Dreiner, Haber, and
  Martin}}]{Dreiner:2008tw}
\bibinfo{author}{\bibfnamefont{H.~K.} \bibnamefont{Dreiner}},
  \bibinfo{author}{\bibfnamefont{H.~E.} \bibnamefont{Haber}}, \bibnamefont{and}
  \bibinfo{author}{\bibfnamefont{S.~P.} \bibnamefont{Martin}}
  (\bibinfo{year}{2008}), \eprint{0812.1594}.

\bibitem[{\citenamefont{Cassel}(2009)}]{Cassel:2009wt}
\bibinfo{author}{\bibfnamefont{S.}~\bibnamefont{Cassel}}
  (\bibinfo{year}{2009}), \eprint{0903.5307}.

\bibitem[{\citenamefont{Slatyer}(2010)}]{Slatyer:2009vg}
\bibinfo{author}{\bibfnamefont{T.~R.} \bibnamefont{Slatyer}},
  \bibinfo{journal}{JCAP} \textbf{\bibinfo{volume}{1002}}, \bibinfo{pages}{028}
  (\bibinfo{year}{2010}), \eprint{0910.5713}.

\bibitem[{\citenamefont{Padmanabhan and Finkbeiner}(2005)}]{Padmanabhan:2005es}
\bibinfo{author}{\bibfnamefont{N.}~\bibnamefont{Padmanabhan}} \bibnamefont{and}
  \bibinfo{author}{\bibfnamefont{D.~P.} \bibnamefont{Finkbeiner}},
  \bibinfo{journal}{Phys. Rev.} \textbf{\bibinfo{volume}{D72}},
  \bibinfo{pages}{023508} (\bibinfo{year}{2005}), \eprint{astro-ph/0503486}.

\bibitem[{\citenamefont{Galli et~al.}(2009)\citenamefont{Galli, Iocco, Bertone,
  and Melchiorri}}]{Galli:2009zc}
\bibinfo{author}{\bibfnamefont{S.}~\bibnamefont{Galli}},
  \bibinfo{author}{\bibfnamefont{F.}~\bibnamefont{Iocco}},
  \bibinfo{author}{\bibfnamefont{G.}~\bibnamefont{Bertone}}, \bibnamefont{and}
  \bibinfo{author}{\bibfnamefont{A.}~\bibnamefont{Melchiorri}},
  \bibinfo{journal}{Phys. Rev.} \textbf{\bibinfo{volume}{D80}},
  \bibinfo{pages}{023505} (\bibinfo{year}{2009}), \eprint{0905.0003}.

\bibitem[{\citenamefont{Slatyer et~al.}(2009)\citenamefont{Slatyer,
  Padmanabhan, and Finkbeiner}}]{Slatyer:2009yq}
\bibinfo{author}{\bibfnamefont{T.~R.} \bibnamefont{Slatyer}},
  \bibinfo{author}{\bibfnamefont{N.}~\bibnamefont{Padmanabhan}},
  \bibnamefont{and} \bibinfo{author}{\bibfnamefont{D.~P.}
  \bibnamefont{Finkbeiner}}, \bibinfo{journal}{Phys. Rev.}
  \textbf{\bibinfo{volume}{D80}}, \bibinfo{pages}{043526}
  (\bibinfo{year}{2009}), \eprint{0906.1197}.

\bibitem[{\citenamefont{Buckley and Fox}(2010)}]{Buckley:2009in}
\bibinfo{author}{\bibfnamefont{M.~R.} \bibnamefont{Buckley}} \bibnamefont{and}
  \bibinfo{author}{\bibfnamefont{P.~J.} \bibnamefont{Fox}},
  \bibinfo{journal}{Phys. Rev.} \textbf{\bibinfo{volume}{D81}},
  \bibinfo{pages}{083522} (\bibinfo{year}{2010}), \eprint{0911.3898}.

\bibitem[{\citenamefont{Feng et~al.}(2009)\citenamefont{Feng, Kaplinghat, and
  Yu}}]{Feng:2009hw}
\bibinfo{author}{\bibfnamefont{J.~L.} \bibnamefont{Feng}},
  \bibinfo{author}{\bibfnamefont{M.}~\bibnamefont{Kaplinghat}},
  \bibnamefont{and} \bibinfo{author}{\bibfnamefont{H.-B.} \bibnamefont{Yu}}
  (\bibinfo{year}{2009}), \eprint{0911.0422}.

\bibitem[{\citenamefont{Khrapak et~al.}(2003)\citenamefont{Khrapak, Ivlev,
  Morfill, and Zhdanov}}]{PhysRevLett.90.225002}
\bibinfo{author}{\bibfnamefont{S.~A.} \bibnamefont{Khrapak}},
  \bibinfo{author}{\bibfnamefont{A.~V.} \bibnamefont{Ivlev}},
  \bibinfo{author}{\bibfnamefont{G.~E.} \bibnamefont{Morfill}},
  \bibnamefont{and} \bibinfo{author}{\bibfnamefont{S.~K.}
  \bibnamefont{Zhdanov}}, \bibinfo{journal}{Phys. Rev. Lett.}
  \textbf{\bibinfo{volume}{90}}, \bibinfo{pages}{225002}
  (\bibinfo{year}{2003}).

\bibitem[{\citenamefont{Hannestad}(2000)}]{Hannestad:2000bs}
\bibinfo{author}{\bibfnamefont{S.}~\bibnamefont{Hannestad}}
  (\bibinfo{year}{2000}), \eprint{astro-ph/0008422}.

\bibitem[{\citenamefont{Loeb and Weiner}(2010)}]{Loeb:2010gj}
\bibinfo{author}{\bibfnamefont{A.}~\bibnamefont{Loeb}} \bibnamefont{and}
  \bibinfo{author}{\bibfnamefont{N.}~\bibnamefont{Weiner}}
  (\bibinfo{year}{2010}), \eprint{1011.6374}.

\bibitem[{\citenamefont{Feng et~al.}(2010)\citenamefont{Feng, Kaplinghat, and
  Yu}}]{Feng:2010zp}
\bibinfo{author}{\bibfnamefont{J.~L.} \bibnamefont{Feng}},
  \bibinfo{author}{\bibfnamefont{M.}~\bibnamefont{Kaplinghat}},
  \bibnamefont{and} \bibinfo{author}{\bibfnamefont{H.-B.} \bibnamefont{Yu}}
  (\bibinfo{year}{2010}), \eprint{1005.4678}.

\bibitem[{\citenamefont{Bell and Jacques}(2008)}]{Bell:2008vx}
\bibinfo{author}{\bibfnamefont{N.~F.} \bibnamefont{Bell}} \bibnamefont{and}
  \bibinfo{author}{\bibfnamefont{T.~D.} \bibnamefont{Jacques}}
  (\bibinfo{year}{2008}), \eprint{0811.0821}.

\bibitem[{\citenamefont{Bertone et~al.}(2008)\citenamefont{Bertone, Cirelli,
  Strumia, and Taoso}}]{Bertone:2008xr}
\bibinfo{author}{\bibfnamefont{G.}~\bibnamefont{Bertone}},
  \bibinfo{author}{\bibfnamefont{M.}~\bibnamefont{Cirelli}},
  \bibinfo{author}{\bibfnamefont{A.}~\bibnamefont{Strumia}}, \bibnamefont{and}
  \bibinfo{author}{\bibfnamefont{M.}~\bibnamefont{Taoso}}
  (\bibinfo{year}{2008}), \eprint{0811.3744}.

\bibitem[{\citenamefont{Bergstrom et~al.}(2008)\citenamefont{Bergstrom,
  Bertone, Bringmann, Edsjo, and Taoso}}]{Bergstrom:2008ag}
\bibinfo{author}{\bibfnamefont{L.}~\bibnamefont{Bergstrom}},
  \bibinfo{author}{\bibfnamefont{G.}~\bibnamefont{Bertone}},
  \bibinfo{author}{\bibfnamefont{T.}~\bibnamefont{Bringmann}},
  \bibinfo{author}{\bibfnamefont{J.}~\bibnamefont{Edsjo}}, \bibnamefont{and}
  \bibinfo{author}{\bibfnamefont{M.}~\bibnamefont{Taoso}}
  (\bibinfo{year}{2008}), \eprint{0812.3895}.

\bibitem[{\citenamefont{Cirelli and Panci}(2009)}]{Cirelli:2009vg}
\bibinfo{author}{\bibfnamefont{M.}~\bibnamefont{Cirelli}} \bibnamefont{and}
  \bibinfo{author}{\bibfnamefont{P.}~\bibnamefont{Panci}},
  \bibinfo{journal}{Nucl. Phys.} \textbf{\bibinfo{volume}{B821}},
  \bibinfo{pages}{399} (\bibinfo{year}{2009}), \eprint{0904.3830}.

\bibitem[{\citenamefont{Pato et~al.}(2009)\citenamefont{Pato, Pieri, and
  Bertone}}]{Pato:2009fn}
\bibinfo{author}{\bibfnamefont{M.}~\bibnamefont{Pato}},
  \bibinfo{author}{\bibfnamefont{L.}~\bibnamefont{Pieri}}, \bibnamefont{and}
  \bibinfo{author}{\bibfnamefont{G.}~\bibnamefont{Bertone}},
  \bibinfo{journal}{Phys. Rev.} \textbf{\bibinfo{volume}{D80}},
  \bibinfo{pages}{103510} (\bibinfo{year}{2009}), \eprint{0905.0372}.

\bibitem[{\citenamefont{Meade et~al.}(2010)\citenamefont{Meade, Papucci,
  Strumia, and Volansky}}]{Meade:2009iu}
\bibinfo{author}{\bibfnamefont{P.}~\bibnamefont{Meade}},
  \bibinfo{author}{\bibfnamefont{M.}~\bibnamefont{Papucci}},
  \bibinfo{author}{\bibfnamefont{A.}~\bibnamefont{Strumia}}, \bibnamefont{and}
  \bibinfo{author}{\bibfnamefont{T.}~\bibnamefont{Volansky}},
  \bibinfo{journal}{Nucl. Phys.} \textbf{\bibinfo{volume}{B831}},
  \bibinfo{pages}{178} (\bibinfo{year}{2010}), \eprint{0905.0480}.

\bibitem[{\citenamefont{Cirelli et~al.}(2010)\citenamefont{Cirelli, Panci, and
  Serpico}}]{Cirelli:2009dv}
\bibinfo{author}{\bibfnamefont{M.}~\bibnamefont{Cirelli}},
  \bibinfo{author}{\bibfnamefont{P.}~\bibnamefont{Panci}}, \bibnamefont{and}
  \bibinfo{author}{\bibfnamefont{P.~D.} \bibnamefont{Serpico}},
  \bibinfo{journal}{Nucl. Phys.} \textbf{\bibinfo{volume}{B840}},
  \bibinfo{pages}{284} (\bibinfo{year}{2010}), \eprint{0912.0663}.

\bibitem[{\citenamefont{Papucci and Strumia}(2010)}]{Papucci:2009gd}
\bibinfo{author}{\bibfnamefont{M.}~\bibnamefont{Papucci}} \bibnamefont{and}
  \bibinfo{author}{\bibfnamefont{A.}~\bibnamefont{Strumia}},
  \bibinfo{journal}{JCAP} \textbf{\bibinfo{volume}{1003}}, \bibinfo{pages}{014}
  (\bibinfo{year}{2010}), \eprint{0912.0742}.

\bibitem[{\citenamefont{Hutsi et~al.}(2010)\citenamefont{Hutsi, Hektor, and
  Raidal}}]{Hutsi:2010ai}
\bibinfo{author}{\bibfnamefont{G.}~\bibnamefont{Hutsi}},
  \bibinfo{author}{\bibfnamefont{A.}~\bibnamefont{Hektor}}, \bibnamefont{and}
  \bibinfo{author}{\bibfnamefont{M.}~\bibnamefont{Raidal}},
  \bibinfo{journal}{JCAP} \textbf{\bibinfo{volume}{1007}}, \bibinfo{pages}{008}
  (\bibinfo{year}{2010}), \eprint{1004.2036}.

\bibitem[{\citenamefont{{de Blok}}(2010)}]{2010AdAst2010E...5D}
\bibinfo{author}{\bibfnamefont{W.~J.~G.} \bibnamefont{{de Blok}}},
  \bibinfo{journal}{Advances in Astronomy} \textbf{\bibinfo{volume}{2010}}
  (\bibinfo{year}{2010}), \eprint{0910.3538}.

\bibitem[{\citenamefont{Oh et~al.}(2010)\citenamefont{Oh, Brook, Governato,
  Brinks, Mayer et~al.}}]{Oh:2010mc}
\bibinfo{author}{\bibfnamefont{S.-H.} \bibnamefont{Oh}},
  \bibinfo{author}{\bibfnamefont{C.}~\bibnamefont{Brook}},
  \bibinfo{author}{\bibfnamefont{F.}~\bibnamefont{Governato}},
  \bibinfo{author}{\bibfnamefont{E.}~\bibnamefont{Brinks}},
  \bibinfo{author}{\bibfnamefont{L.}~\bibnamefont{Mayer}}, \bibnamefont{et~al.}
  (\bibinfo{year}{2010}), \eprint{1011.2777}.

\bibitem[{\citenamefont{Zaharijas et~al.}(2010)\citenamefont{Zaharijas, Cuoco,
  Yang, and Conrad}}]{Zaharijas:2010ca}
\bibinfo{author}{\bibfnamefont{G.}~\bibnamefont{Zaharijas}},
  \bibinfo{author}{\bibfnamefont{A.}~\bibnamefont{Cuoco}},
  \bibinfo{author}{\bibfnamefont{Z.}~\bibnamefont{Yang}}, \bibnamefont{and}
  \bibinfo{author}{\bibfnamefont{J.}~\bibnamefont{Conrad}}
  (\bibinfo{collaboration}{for the Fermi-LAT}) (\bibinfo{year}{2010}),
  \eprint{1012.0588}.

\bibitem[{\citenamefont{Cirelli and Cline}(2010)}]{Cirelli:2010nh}
\bibinfo{author}{\bibfnamefont{M.}~\bibnamefont{Cirelli}} \bibnamefont{and}
  \bibinfo{author}{\bibfnamefont{J.~M.} \bibnamefont{Cline}}
  (\bibinfo{year}{2010}), \eprint{1005.1779}.

\bibitem[{\citenamefont{Navarro et~al.}(2008)}]{Navarro:2008kc}
\bibinfo{author}{\bibfnamefont{J.~F.} \bibnamefont{Navarro}}
  \bibnamefont{et~al.} (\bibinfo{year}{2008}), \eprint{0810.1522}.

\bibitem[{\citenamefont{{Blumenthal} et~al.}(1986)\citenamefont{{Blumenthal},
  {Faber}, {Flores}, and {Primack}}}]{1986ApJ...301...27B}
\bibinfo{author}{\bibfnamefont{G.~R.} \bibnamefont{{Blumenthal}}},
  \bibinfo{author}{\bibfnamefont{S.~M.} \bibnamefont{{Faber}}},
  \bibinfo{author}{\bibfnamefont{R.}~\bibnamefont{{Flores}}}, \bibnamefont{and}
  \bibinfo{author}{\bibfnamefont{J.~R.} \bibnamefont{{Primack}}},
  \bibinfo{journal}{\apj} \textbf{\bibinfo{volume}{301}}, \bibinfo{pages}{27}
  (\bibinfo{year}{1986}).

\bibitem[{\citenamefont{{Gnedin} et~al.}(2004)\citenamefont{{Gnedin},
  {Kravtsov}, {Klypin}, and {Nagai}}}]{2004ApJ...616...16G}
\bibinfo{author}{\bibfnamefont{O.~Y.} \bibnamefont{{Gnedin}}},
  \bibinfo{author}{\bibfnamefont{A.~V.} \bibnamefont{{Kravtsov}}},
  \bibinfo{author}{\bibfnamefont{A.~A.} \bibnamefont{{Klypin}}},
  \bibnamefont{and} \bibinfo{author}{\bibfnamefont{D.}~\bibnamefont{{Nagai}}},
  \bibinfo{journal}{\apj} \textbf{\bibinfo{volume}{616}}, \bibinfo{pages}{16}
  (\bibinfo{year}{2004}), \eprint{arXiv:astro-ph/0406247}.

\bibitem[{\citenamefont{El-Zant et~al.}(2001)\citenamefont{El-Zant, Shlosman,
  and Hoffman}}]{ElZant:2001re}
\bibinfo{author}{\bibfnamefont{A.}~\bibnamefont{El-Zant}},
  \bibinfo{author}{\bibfnamefont{I.}~\bibnamefont{Shlosman}}, \bibnamefont{and}
  \bibinfo{author}{\bibfnamefont{Y.}~\bibnamefont{Hoffman}}
  (\bibinfo{year}{2001}), \eprint{astro-ph/0103386}.

\bibitem[{\citenamefont{El-Zant et~al.}(2004)\citenamefont{El-Zant, Hoffman,
  Primack, Combes, and Shlosman}}]{ElZant:2003rp}
\bibinfo{author}{\bibfnamefont{A.~A.} \bibnamefont{El-Zant}},
  \bibinfo{author}{\bibfnamefont{Y.}~\bibnamefont{Hoffman}},
  \bibinfo{author}{\bibfnamefont{J.}~\bibnamefont{Primack}},
  \bibinfo{author}{\bibfnamefont{F.}~\bibnamefont{Combes}}, \bibnamefont{and}
  \bibinfo{author}{\bibfnamefont{I.}~\bibnamefont{Shlosman}},
  \bibinfo{journal}{Astrophys. J.} \textbf{\bibinfo{volume}{607}},
  \bibinfo{pages}{L75} (\bibinfo{year}{2004}), \eprint{astro-ph/0309412}.

\bibitem[{\citenamefont{Romano-Diaz et~al.}(2008)\citenamefont{Romano-Diaz,
  Shlosman, Hoffman, and Heller}}]{RomanoDiaz:2008wz}
\bibinfo{author}{\bibfnamefont{E.}~\bibnamefont{Romano-Diaz}},
  \bibinfo{author}{\bibfnamefont{I.}~\bibnamefont{Shlosman}},
  \bibinfo{author}{\bibfnamefont{Y.}~\bibnamefont{Hoffman}}, \bibnamefont{and}
  \bibinfo{author}{\bibfnamefont{C.}~\bibnamefont{Heller}}
  (\bibinfo{year}{2008}), \eprint{0808.0195}.

\bibitem[{\citenamefont{Governato et~al.}(2009)}]{Governato:2009bg}
\bibinfo{author}{\bibfnamefont{F.}~\bibnamefont{Governato}}
  \bibnamefont{et~al.} (\bibinfo{year}{2009}), \eprint{0911.2237}.

\bibitem[{\citenamefont{Romano-Diaz et~al.}(2009)\citenamefont{Romano-Diaz,
  Shlosman, Heller, and Hoffman}}]{RomanoDiaz:2009yq}
\bibinfo{author}{\bibfnamefont{E.}~\bibnamefont{Romano-Diaz}},
  \bibinfo{author}{\bibfnamefont{I.}~\bibnamefont{Shlosman}},
  \bibinfo{author}{\bibfnamefont{C.}~\bibnamefont{Heller}}, \bibnamefont{and}
  \bibinfo{author}{\bibfnamefont{Y.}~\bibnamefont{Hoffman}},
  \bibinfo{journal}{Astrophys. J.} \textbf{\bibinfo{volume}{702}},
  \bibinfo{pages}{1250} (\bibinfo{year}{2009}), \eprint{0901.1317}.

\bibitem[{\citenamefont{Abadi et~al.}(2009)\citenamefont{Abadi, Navarro,
  Fardal, Babul, and Steinmetz}}]{Abadi:2009ve}
\bibinfo{author}{\bibfnamefont{M.~G.} \bibnamefont{Abadi}},
  \bibinfo{author}{\bibfnamefont{J.~F.} \bibnamefont{Navarro}},
  \bibinfo{author}{\bibfnamefont{M.}~\bibnamefont{Fardal}},
  \bibinfo{author}{\bibfnamefont{A.}~\bibnamefont{Babul}}, \bibnamefont{and}
  \bibinfo{author}{\bibfnamefont{M.}~\bibnamefont{Steinmetz}}
  (\bibinfo{year}{2009}), \eprint{0902.2477}.

\bibitem[{\citenamefont{Pedrosa et~al.}(2009)\citenamefont{Pedrosa, Tissera,
  and Scannapieco}}]{Pedrosa:2009rw}
\bibinfo{author}{\bibfnamefont{S.~E.} \bibnamefont{Pedrosa}},
  \bibinfo{author}{\bibfnamefont{P.~B.} \bibnamefont{Tissera}},
  \bibnamefont{and}
  \bibinfo{author}{\bibfnamefont{C.}~\bibnamefont{Scannapieco}}
  (\bibinfo{year}{2009}), \eprint{0910.4380}.

\bibitem[{\citenamefont{Tissera et~al.}(2009)\citenamefont{Tissera, White,
  Pedrosa, and Scannapieco}}]{Tissera:2009cm}
\bibinfo{author}{\bibfnamefont{P.~B.} \bibnamefont{Tissera}},
  \bibinfo{author}{\bibfnamefont{S.~D.~M.} \bibnamefont{White}},
  \bibinfo{author}{\bibfnamefont{S.}~\bibnamefont{Pedrosa}}, \bibnamefont{and}
  \bibinfo{author}{\bibfnamefont{C.}~\bibnamefont{Scannapieco}}
  (\bibinfo{year}{2009}), \eprint{0911.2316}.

\bibitem[{\citenamefont{Su et~al.}(2010)\citenamefont{Su, Slatyer, and
  Finkbeiner}}]{Su:2010qj}
\bibinfo{author}{\bibfnamefont{M.}~\bibnamefont{Su}},
  \bibinfo{author}{\bibfnamefont{T.~R.} \bibnamefont{Slatyer}},
  \bibnamefont{and} \bibinfo{author}{\bibfnamefont{D.~P.}
  \bibnamefont{Finkbeiner}} (\bibinfo{year}{2010}), \eprint{1005.5480}.

\bibitem[{\citenamefont{Delahaye et~al.}(2009)}]{Delahaye:2008ua}
\bibinfo{author}{\bibfnamefont{T.}~\bibnamefont{Delahaye}}
  \bibnamefont{et~al.}, \bibinfo{journal}{Astron. Astrophys.}
  \textbf{\bibinfo{volume}{501}}, \bibinfo{pages}{821} (\bibinfo{year}{2009}),
  \eprint{0809.5268}.

\bibitem[{\citenamefont{Dent et~al.}(2010)\citenamefont{Dent, Dutta, and
  Scherrer}}]{Dent:2009bv}
\bibinfo{author}{\bibfnamefont{J.~B.} \bibnamefont{Dent}},
  \bibinfo{author}{\bibfnamefont{S.}~\bibnamefont{Dutta}}, \bibnamefont{and}
  \bibinfo{author}{\bibfnamefont{R.~J.} \bibnamefont{Scherrer}},
  \bibinfo{journal}{Phys. Lett.} \textbf{\bibinfo{volume}{B687}},
  \bibinfo{pages}{275} (\bibinfo{year}{2010}), \eprint{0909.4128}.

\bibitem[{\citenamefont{Zavala et~al.}(2010{\natexlab{a}})\citenamefont{Zavala,
  Vogelsberger, and White}}]{Zavala:2009mi}
\bibinfo{author}{\bibfnamefont{J.}~\bibnamefont{Zavala}},
  \bibinfo{author}{\bibfnamefont{M.}~\bibnamefont{Vogelsberger}},
  \bibnamefont{and} \bibinfo{author}{\bibfnamefont{S.~D.~M.}
  \bibnamefont{White}}, \bibinfo{journal}{Phys. Rev.}
  \textbf{\bibinfo{volume}{D81}}, \bibinfo{pages}{083502}
  (\bibinfo{year}{2010}{\natexlab{a}}), \eprint{0910.5221}.

\bibitem[{\citenamefont{Essig et~al.}(2010)\citenamefont{Essig, Sehgal,
  Strigari, Geha, and Simon}}]{Essig:2010em}
\bibinfo{author}{\bibfnamefont{R.}~\bibnamefont{Essig}},
  \bibinfo{author}{\bibfnamefont{N.}~\bibnamefont{Sehgal}},
  \bibinfo{author}{\bibfnamefont{L.~E.} \bibnamefont{Strigari}},
  \bibinfo{author}{\bibfnamefont{M.}~\bibnamefont{Geha}}, \bibnamefont{and}
  \bibinfo{author}{\bibfnamefont{J.~D.} \bibnamefont{Simon}}
  (\bibinfo{year}{2010}), \eprint{1007.4199}.

\bibitem[{\citenamefont{Mardon et~al.}(2009)\citenamefont{Mardon, Nomura,
  Stolarski, and Thaler}}]{Mardon:2009rc}
\bibinfo{author}{\bibfnamefont{J.}~\bibnamefont{Mardon}},
  \bibinfo{author}{\bibfnamefont{Y.}~\bibnamefont{Nomura}},
  \bibinfo{author}{\bibfnamefont{D.}~\bibnamefont{Stolarski}},
  \bibnamefont{and} \bibinfo{author}{\bibfnamefont{J.}~\bibnamefont{Thaler}},
  \bibinfo{journal}{JCAP} \textbf{\bibinfo{volume}{0905}}, \bibinfo{pages}{016}
  (\bibinfo{year}{2009}), \eprint{0901.2926}.

\bibitem[{\citenamefont{Pieri et~al.}(2011)\citenamefont{Pieri, Lavalle,
  Bertone, and Branchini}}]{Pieri:2009je}
\bibinfo{author}{\bibfnamefont{L.}~\bibnamefont{Pieri}},
  \bibinfo{author}{\bibfnamefont{J.}~\bibnamefont{Lavalle}},
  \bibinfo{author}{\bibfnamefont{G.}~\bibnamefont{Bertone}}, \bibnamefont{and}
  \bibinfo{author}{\bibfnamefont{E.}~\bibnamefont{Branchini}},
  \bibinfo{journal}{Phys. Rev.} \textbf{\bibinfo{volume}{D83}},
  \bibinfo{pages}{023518} (\bibinfo{year}{2011}), \eprint{0908.0195}.

\bibitem[{\citenamefont{Abazajian et~al.}(2010)\citenamefont{Abazajian,
  Blanchet, and Harding}}]{Abazajian:2010zb}
\bibinfo{author}{\bibfnamefont{K.~N.} \bibnamefont{Abazajian}},
  \bibinfo{author}{\bibfnamefont{S.}~\bibnamefont{Blanchet}}, \bibnamefont{and}
  \bibinfo{author}{\bibfnamefont{J.~P.} \bibnamefont{Harding}}
  (\bibinfo{year}{2010}), \eprint{1011.5090}.

\bibitem[{\citenamefont{Abdo et~al.}(2010)}]{Abdo:2010dk}
\bibinfo{author}{\bibfnamefont{A.~A.} \bibnamefont{Abdo}} \bibnamefont{et~al.}
  (\bibinfo{collaboration}{Fermi-LAT}), \bibinfo{journal}{JCAP}
  \textbf{\bibinfo{volume}{1004}}, \bibinfo{pages}{014} (\bibinfo{year}{2010}),
  \eprint{1002.4415}.

\bibitem[{\citenamefont{Zavala et~al.}(2011)\citenamefont{Zavala, Vogelsberger,
  Slatyer, Loeb, and Springel}}]{Zavala:2011tt}
\bibinfo{author}{\bibfnamefont{J.}~\bibnamefont{Zavala}},
  \bibinfo{author}{\bibfnamefont{M.}~\bibnamefont{Vogelsberger}},
  \bibinfo{author}{\bibfnamefont{T.~R.} \bibnamefont{Slatyer}},
  \bibinfo{author}{\bibfnamefont{A.}~\bibnamefont{Loeb}}, \bibnamefont{and}
  \bibinfo{author}{\bibfnamefont{V.}~\bibnamefont{Springel}}
  (\bibinfo{year}{2011}), \eprint{1103.0776}.

\bibitem[{\citenamefont{Zavala et~al.}(2010{\natexlab{b}})\citenamefont{Zavala,
  Springel, and Boylan-Kolchin}}]{Zavala:2009zr}
\bibinfo{author}{\bibfnamefont{J.}~\bibnamefont{Zavala}},
  \bibinfo{author}{\bibfnamefont{V.}~\bibnamefont{Springel}}, \bibnamefont{and}
  \bibinfo{author}{\bibfnamefont{M.}~\bibnamefont{Boylan-Kolchin}},
  \bibinfo{journal}{Mon. Not. Roy. Astron. Soc.}
  \textbf{\bibinfo{volume}{405}}, \bibinfo{pages}{593}
  (\bibinfo{year}{2010}{\natexlab{b}}), \eprint{0908.2428}.

\bibitem[{\citenamefont{Bringmann}(2009)}]{Bringmann:2009vf}
\bibinfo{author}{\bibfnamefont{T.}~\bibnamefont{Bringmann}},
  \bibinfo{journal}{New J. Phys.} \textbf{\bibinfo{volume}{11}},
  \bibinfo{pages}{105027} (\bibinfo{year}{2009}), \eprint{0903.0189}.

\bibitem[{\citenamefont{Hisano et~al.}(2004)\citenamefont{Hisano, Matsumoto,
  and Nojiri}}]{Hisano:2003ec}
\bibinfo{author}{\bibfnamefont{J.}~\bibnamefont{Hisano}},
  \bibinfo{author}{\bibfnamefont{S.}~\bibnamefont{Matsumoto}},
  \bibnamefont{and} \bibinfo{author}{\bibfnamefont{M.~M.}
  \bibnamefont{Nojiri}}, \bibinfo{journal}{Phys. Rev. Lett.}
  \textbf{\bibinfo{volume}{92}}, \bibinfo{pages}{031303}
  (\bibinfo{year}{2004}), \eprint{hep-ph/0307216}.

\bibitem[{\citenamefont{Hisano et~al.}(2005)\citenamefont{Hisano, Matsumoto,
  Nojiri, and Saito}}]{Hisano:2004ds}
\bibinfo{author}{\bibfnamefont{J.}~\bibnamefont{Hisano}},
  \bibinfo{author}{\bibfnamefont{S.}~\bibnamefont{Matsumoto}},
  \bibinfo{author}{\bibfnamefont{M.~M.} \bibnamefont{Nojiri}},
  \bibnamefont{and} \bibinfo{author}{\bibfnamefont{O.}~\bibnamefont{Saito}},
  \bibinfo{journal}{Phys. Rev.} \textbf{\bibinfo{volume}{D71}},
  \bibinfo{pages}{063528} (\bibinfo{year}{2005}), \eprint{hep-ph/0412403}.

\bibitem[{\citenamefont{Galli et~al.}(2011)\citenamefont{Galli, Iocco, Bertone,
  and Melchiorri}}]{Galli:2011rz}
\bibinfo{author}{\bibfnamefont{S.}~\bibnamefont{Galli}},
  \bibinfo{author}{\bibfnamefont{F.}~\bibnamefont{Iocco}},
  \bibinfo{author}{\bibfnamefont{G.}~\bibnamefont{Bertone}}, \bibnamefont{and}
  \bibinfo{author}{\bibfnamefont{A.}~\bibnamefont{Melchiorri}}
  (\bibinfo{year}{2011}), \eprint{1106.1528}.

\bibitem[{\citenamefont{Hutsi et~al.}(2011)\citenamefont{Hutsi, Chluba, Hektor,
  and Raidal}}]{Hutsi:2011vx}
\bibinfo{author}{\bibfnamefont{G.}~\bibnamefont{Hutsi}},
  \bibinfo{author}{\bibfnamefont{J.}~\bibnamefont{Chluba}},
  \bibinfo{author}{\bibfnamefont{A.}~\bibnamefont{Hektor}}, \bibnamefont{and}
  \bibinfo{author}{\bibfnamefont{M.}~\bibnamefont{Raidal}}
  (\bibinfo{year}{2011}), \eprint{1103.2766}.

\end{thebibliography}
\bibliographystyle{apsrev}

\end{document}